\RequirePackage{ifluatex}

\documentclass{article}

\usepackage{arxiv}

\usepackage[utf8]{inputenc} % allow utf-8 input
\usepackage[T1]{fontenc}    % use 8-bit T1 fonts
\usepackage{hyperref}       % hyperlinks
\usepackage{url}            % simple URL typesetting
\usepackage{booktabs}       % professional-quality tables
\usepackage{amsfonts}       % blackboard math symbols
\usepackage{nicefrac}       % compact symbols for 1/2, etc.
\usepackage{microtype}      % microtypography
\usepackage{lipsum}		% Can be removed after putting your text content
\usepackage{graphicx}
\usepackage{doi}

\usepackage{xcolor}
\usepackage{amsmath}

\makeatletter
\newcommand*{\rom}[1]{\expandafter\@slowromancap\romannumeral #1@}

\usepackage{xargs}
\usepackage{geometry}
\usepackage{fancyhdr} % For custom headers
\usepackage{lastpage} % To determine the last page for the footer
\usepackage{extramarks} % For headers and footers
\usepackage[most]{tcolorbox} % For problem answer sections
\usepackage{graphicx} % For inserting images
\usepackage{xcolor} % For link coloring

\usepackage[export]{adjustbox}
\usepackage[parfill]{parskip}
\usepackage{amssymb}
\newlength{\minuslength}
\usepackage{algorithm}
\usepackage{algorithmic}
\usepackage{caption} 
\usepackage{subcaption}
\usepackage{multirow}
\usepackage{setspace}
\title{Approximate Bayesian Inference for the Interaction Types \rom{1},\rom{2}, \rom{3} and \rom{4} with Application in Disease Mapping}

\author{ \href{https://orcid.org/0000-0003-1587-3288}{\includegraphics[scale=0.06]{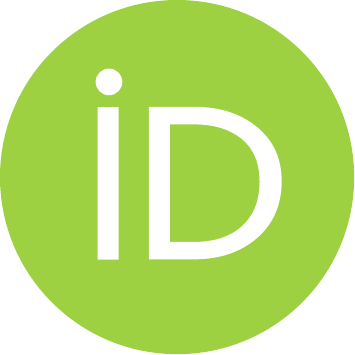}\hspace{1mm}Esmail Abdul Fattah}\thanks{Corresponding Author} \\
	Statistics Program, CEMSE Division\\
	King Abdullah University of Science and Technology\\
	Thuwal, 23955, Makkah\\
	\texttt{esmail.abdulfattah@kaust.edu.sa} \\
	\And
	\href{https://orcid.org/0000-0002-0222-1881}H{\aa}vard Rue \\
	Statistics Program, CEMSE Division\\
	King Abdullah University of Science and Technology\\
	Thuwal, 23955, Makkah\\
	\texttt{haavard.rue@kaust.edu.sa} \\
}

\renewcommand{\shorttitle}{Approximate Bayesian Inference for the Interaction Types \rom{1}, \rom{2}, \rom{3}, and \rom{4}}

\begin{document}
\maketitle

\begin{abstract}

We address in this paper a new approach for fitting spatiotemporal models with application in disease mapping using the interaction types \rom{1}, \rom{2}, \rom{3}, and \rom{4} proposed by \cite{KnorrHeld2000BayesianMO}. When we account for the spatiotemporal interactions in disease-mapping models, inference becomes more useful in revealing unknown patterns in the data. However, when the number of locations and/or the number of time points is large, the inference gets computationally challenging due to the high number of required constraints necessary for inference, and this holds for various inference architectures including Markov chain Monte Carlo (MCMC) \cite{Gilks1997MarkovCM} and Integrated Nested Laplace Approximations (INLA) \cite{Rue2009ApproximateBI}. We re-formulate INLA approach based on dense matrices to fit the intrinsic spatiotemporal models with the four interaction types and account for the sum-to-zero constraints, and discuss how the new approach can be implemented in a high-performance computing framework. The computing time using the new approach does not depend on the number of constraints and can reach a 40-fold faster speed compared to INLA in realistic scenarios. This approach is verified by a simulation study and a real data application, and it is implemented in the R package \href{https://github.com/esmail-abdulfattah/INLAPLUS}{\texttt{INLAPLUS}}  and the Python header function: \href{https://github.com/esmail-abdulfattah/INLAPLUS}{\texttt{inla1234()}}.
\end{abstract}

% keywords can be removed

\section{Introduction}
Spatiotemporal disease-mapping models have become increasingly common across different disciplines and application areas, including epidemiology, public health, and medicine. This is partly due to the recent unprecedented availability of data production and resources, and the need to describe the variations in risk over time and space. The analyzed outcome of spatiotemporal models depends on the data collected across both time and space, and these models have at least one spatial and one temporal effect. This requires taking heed of spatial correlations and temporal correlations. Exploiting both interrelations adds complexity to the model due to the influence of the neighboring structure, as the significant spatiotemporal interactions should be also given consideration. For instance, meteorologists are interested in modeling rainfall intermittency, which depends on the temporal and spatial structure as well as the spatiotemporal interactions. This consideration also applies when modeling a spatiotemporal trend of COVID-19 risk in Malawi \cite{Ngwira2021SpatialTD}. 

Several proposed formulations exist for the spatiotemporal interaction types including \cite{Waller1997HierarchicalSM,Lagazio2003AgeperiodcohortMA, Schmid2004BayesianEO}. \cite{KnorrHeld2000BayesianMO} proposes the four flexible interaction types \rom{1}, \rom{2}, \rom{3}, and \rom{4} for space-time that do not need to follow any specific trend. The design of the precision matrix for the spatiotemporal interactions is the Kronecker product of a spatial random effect by a time random effect. Each formulation of these interaction types is a combination of structured and unstructured effects. These types are still commonly adopted in many applications: a crash frequency model for improving safety analytics \cite{Cui2021AnAH}, online bus speed prediction \cite{Cui2021OnlineBS}, dynamic shrinkage process for population shifts \cite{Lym2021ExploringDP}, and associations between social vulnerability, environmental measurements and Covid-19 \cite{Johnson2021SpatiotemporalAB}. We focus in this paper on models using the four interaction types proposed by \cite{KnorrHeld2000BayesianMO}.

Disease-mapping models are usually formulated as generalized linear mixed models (GLMM) within a three-level hierarchical framework using a Poisson or binomial likelihood. One assumption in these four types is that the parameters of the resulting model follow a Gaussian Markov Random Field (GMRF) \cite{Rue2005GaussianMR}. By construction, the models \rom{2}, \rom{3}, and \rom{4} are intrinsic, and constraints are added to make the model identifiable \cite{Rue2005GaussianMR}. These constraints are equivalent to the eigenvectors that span the null space of the deficient precision matrix. To avoid these constraints, the prior model for the interaction terms can be considered as a limit of a proper distribution \cite{MartnezBeneito2008AnAA,Rushworth2014ASM}. However, the main temporal and spatial effects confound with the interaction terms, and this makes interpretations difficult for parameter estimations. The number of constraints depends on the rank of the full precision matrix of the latent field.

The dependency structure of the effects becomes more complex when the number of constraints increases, and this can create a computational bottleneck. This is the case when the main random effects are improper, and the resulting precision matrix of the spatiotemporal interactions is the Kronecker product of at least two improper models, leading to a highly rank-deficient precision matrix. 

Two novel approaches that have been effectively used for fitting spatiotemporal disease-mapping models are Markov chain Monte Carlo (MCMC) \cite{Gilks1997MarkovCM,Brooks2011HandbookOM} and Integrated Nested Laplace Approximations (INLA) \cite{Rue2009ApproximateBI, Rue2016BayesianCW}. They have a wide range of applications: \cite{Schrdle2011SpatiotemporalDM, lawson2013bayesian, Liu2018UsingTM, Hu2018UrbanCP, Sun2021SpatiotemporalMO}, amongst others. MCMC methods still face difficulties in fitting complex spatiotemporal disease-mapping models due to the dimensional and dependency structures, which make them computationally expensive due to the need for numerous iterations \cite{Cui2021AnAH, Lord2010TheSA}, and the high number of constraints that needs to be added to make the model identifiable.

INLA focuses on a specific class of models, Latent Gaussian Models (LGM), which gives it computational advantages and allows fast and accurate model fitting. \cite{Schrdle2011SpatiotemporalDM} show that a wide range of spatiotemporal models can be fitted using INLA. However, LGMs are limited to a certain structure between the response mean and the linear additive predictor. This restricts the use of more complicated models including the presence of realistic dependence structures of the spatiotemporal interaction terms as well as more complex likelihoods, as shown by \cite{Stringer2020ApproximateBI}. The computational complexity for disease-mapping models in INLA depends on the number of constraints $k$ added and becomes expensive for complex interactions due to the strong assumption of sparsity in INLA.

In these intrinsic models, INLA tries to preserve the sparsity of any improper precision matrix to compute approximations by adding a tiny noise to the diagonal, then corrects for constraints using the kriging technique \cite{Rue2005GaussianMR, Schrdle2011SpatiotemporalDM} . The cost when using this technique grows quadratically, and its time complexity is proportional to $\mathcal{O}(sk^2)$, where $s$ is the size of the fixed and random effects. For high $k$, the cost of this technique dominates the overall cost for an approximate inference.
 
Alternatively, instead of correcting for the constraints by dealing with the zero eigenvalues of the improper precision matrices and keeping them sparse, we can work with the covariance matrices and correct for the constraints using the pseudo-inverse. For example, consider a random effect $\pmb \alpha = (\alpha_1, \ldots, \alpha_{6})$ that has a precision structure matrix $\pmb R \in \mathbb{R}^{6 \times 6}$ of rank $5$ and $\pmb R^+$ is its (Moore-Penrose) pseudo-inverse,
\begin{equation*}
  \pmb R = \left(
  \begin{array}{llllll}
 \hspace{\minuslength}1 & -1 &~~~. &~~~. &~~~. &~~~.\\
 -1 & \hspace{\minuslength}2 & -1 &~~~. &~~~. &~~~.\\
 ~~~. & -1 & \hspace{\minuslength}2 & -1 &~~~. &~~~.\\
 ~~~. &~~~. & -1 & \hspace{\minuslength}2 & -1 &~~~.\\
 ~~~. &~~~. &~~~. & -1 & \hspace{\minuslength}2 & -1\\
 ~~~. &~~~. &~~~. &~~~. & -1 & \hspace{\minuslength}1
  \end{array} 
  \right) \text{and~~} \pmb R^{+} = \left( \begin{array}{llllll}
 \hspace{\minuslength}1.53 & \hspace{\minuslength}0.69 & \hspace{\minuslength}0.03 &-0.47 &-0.81 &-0.97\\
 \hspace{\minuslength}0.69 & \hspace{\minuslength}0.86 & \hspace{\minuslength}0.19 &-0.31 &-0.64 &-0.81\\
 \hspace{\minuslength}0.03 & \hspace{\minuslength}0.19 & \hspace{\minuslength}0.53 & \hspace{\minuslength}0.03 &-0.31 &-0.47\\
-0.47 &-0.31 & \hspace{\minuslength}0.03 & \hspace{\minuslength}0.53 & \hspace{\minuslength}0.19 & \hspace{\minuslength}0.03\\
-0.81 &-0.64 &-0.31 & \hspace{\minuslength}0.19 & \hspace{\minuslength}0.86 & \hspace{\minuslength}0.69\\
-0.97 &-0.81 &-0.47 & \hspace{\minuslength}0.03 & \hspace{\minuslength}0.69 & \hspace{\minuslength}1.53
  \end{array} \right).
\end{equation*}

INLA uses the precision structure $\pmb R$ + $\epsilon \pmb I$ (sparse) to compute approximations, where $\epsilon$ is a tiny noise to make the matrix proper, then corrects for constraints by conditioning on $\sum_i \alpha_i = 0$. INLA approach implicitly assumes the number of constraints is small, but this is not always the case. For more complex interactions, we can use the covariance structure, and in our example $\pmb R^+$ (dense) satisfies the condition $\sum_i \alpha_i = 0$.

The sparse matrix assumption is the main computational advantage in INLA methodology, but the use of sparse algorithms is complicated and they need careful implementation. For complex applications, adding more resources to speed up INLA does not scale well using sparse solvers. The parallelization schemes in INLA are based on OpenMP \cite{GaedkeMerzhuser2022ParallelizedIN} using the PARDISO library. This works efficiently on shared-memory architectures, but it cannot scale well on distributed-memory for big applications or for dense structures.  However, we can take the advantage of the multi-core architectures and the abundance of memory in today's computational resources to design a new approach, based on dense matrices, that scales well and utilizes the presence of multiprocessors on shared and distributed memory.

This paper proposes a new approach to approximate Bayesian inference for the four interaction types \rom{1}, \rom{2}, \rom{3}, and \rom{4}. The idea is to re-implement INLA methodology using dense matrices that employ any complex structure for precision/covariance matrices and allow the user to account for constraints by just providing the rank-deficient of each structured effect. In intrinsic models, the precision matrix of the effects is updated using the pseudo-inverse, which makes fitting any spatiotemporal model independent of the number of constraints. The computations are designed to work on both shared and distributed memory to utilize the available computational resources.
 
The rest of this paper is organized as follows. Section 2 presents a general review of the four interaction types presented by \cite{KnorrHeld2000BayesianMO}. Section 3 introduces the proposed approach for fitting spatiotemporal models with interaction types \rom{1}, \rom{2}, \rom{3}, and \rom{4}. Then we demonstrate the use of the new approach through a simulation study and real data application, and we end with a conclusion.

\section{Interaction Types for Spatiotemporal Models} \label{interactiontypes1234}

In this section, we revisit the interaction types proposed by \cite{KnorrHeld2000BayesianMO} to fit non-parametric spatiotemporal models.
We assume the response $\pmb y = (y_1, \ldots, y_{d})$ is associated with a likelihood, including one of the exponential family distributions. In a generalized linear form, we use the linear predictor $\pmb \eta \in \mathbb{R}^{nm}$ to embrace all four different interaction types \rom{1}, \rom{2}, \rom{3}, and \rom{4} \cite{KnorrHeld2000BayesianMO}. The linear predictor accounts additively for the temporal effects, spatial effects, and the spatiotemporal interactions, and is linked to the expected value of the response $\pmb y$ through a link function $g(.)$, such that E$(\pmb y) =  g^{-1}(\pmb \eta)$. Given a set of covariates $\pmb z_{..,1}, \ldots, \pmb z_{..,K} $, we define the form of the linear predictor,
\begin{equation}
    \eta_{ij} = \mu + \displaystyle \sum_{k}^{K} \beta_k z_{ij,k} + \alpha_i + \gamma_i + \delta_j + \phi_j + \varepsilon_{ij}, \quad i = 1, \ldots n, \quad j = 1, \ldots m
    \label{linearpredtype1234}
\end{equation}
\noindent where $\mu$ is the overall intercept, $\beta_1, \ldots, \beta_K$  are the fixed effects of the $K$ covariates, $\alpha_i$ and $\gamma_i$ are the random temporal structured and unstructured effects, $\delta_j$ and $\phi_j$ are the random spatial structured and unstructured effects, and $\varepsilon_{ij}$ represents the interaction terms. The latent field $\pmb x$ is the combination of these fixed and random effects, and in a vector form, $\pmb x = (\mu, \pmb \beta, \pmb \alpha, \pmb \gamma,\pmb \delta,\pmb \phi,\pmb \varepsilon) \in \mathbb{R}^{s}$, such that $s = 1 + K + 2n + 2m + nm$. Each of these effects is assumed to be multivariate Gaussian with mean zero and precision matrix $\pmb Q_{*} = \tau_{*}\pmb R_{*}$, where $* \in \{\alpha, \gamma, \delta, \phi,\varepsilon\}$. We also assume that
\begin{center}
    $\pmb x \sim \mathcal{N}(\pmb 0,\pmb Q_x^{-1})$, $\quad \mu \sim \mathcal{N}(0, \tau_\mu^{-1}) $, $\quad \pmb \beta \sim \mathcal{N}(\pmb 0, \pmb Q_\beta^{-1}),$
\end{center}
where $\pmb Q_x$ is the precision matrix with a known structure but with an unknown set of precision parameters $\pmb \tau = (\tau_\alpha,\tau_\gamma,\tau_\delta,\tau_\phi, \tau_\varepsilon)$ to be estimated from the data, and some block matrices including the interaction matrix $\pmb R_\varepsilon$,
\begin{equation}
    \pmb Q_x = \text{\text{block.diag}}(\tau_\mu, \pmb Q_\beta, \tau_\alpha \pmb R_\alpha,\tau_\gamma \pmb R_\gamma,\tau_\delta \pmb R_\delta,\tau_\phi \pmb R_\phi,\tau_\varepsilon \pmb R_\varepsilon).
    \label{formblockdiagonal}
\end{equation}

Next, we present the different formulations for the unstructured effects, structured effects, and interaction types for space and time.

\subsection{Unstructured Effects on Time and Space}

The unstructured components $\pmb \gamma$ and $\pmb \phi$ are modeled as independent and identically distributed multivariate Gaussian variables (iid) with zero means and covariance matrices $\pmb I_n$ and $\pmb I_m$, respectively.

\subsubsection{Structured Effects: Random Walk and Intrinsic Auto-Regressive Models}

Random Walk and Intrinsic Auto-Regressive (model = "besag" in \texttt{R-INLA}) Models are specified through a neighborhood structure, where every point conditionally depends on some neighbors, as nearby points tend to be similar. Part of the linear predictor in the model decomposes additively of these two structured effects: Random Walk as a temporal effect and Besag as the spatial effect.  

\vspace{0.4cm}
\noindent \underline{\textbf{Random Walk Models}}

\noindent In the four interaction types, we consider $\pmb \alpha$ to be a random walk model of order $l$ = 1 or 2, RW$l$ and have precision matrix $\pmb Q_{\alpha}$ with $\tau_\alpha$ as the precision parameter. We consider the neighboring points tend to be alike with independent Gaussian increments and have equidistant time locations,
\begin{itemize}
    \item Random Walk Model of order 1 (RW1):
    \begin{equation}
        \pi(\pmb \alpha|\tau_\alpha) \propto \exp\Big(-\dfrac{\tau_\alpha}{2} \displaystyle \sum_{i=2}^{n}(\alpha_i - \alpha_{i-1})^2\Big).
    \end{equation}
    \item Random Walk Model of order 2 (RW2):
    \begin{equation}
        \pi(\pmb \alpha|\tau_\alpha) \propto \exp\Big(-\dfrac{\tau_\alpha}{2} \displaystyle \sum_{i=1}^{n-2}(\alpha_i - 2\alpha_{i+1} + \alpha_{i+2})^2\Big).
    \end{equation}
\end{itemize}

\begin{figure}[hbt!]
\centering
\begin{subfigure}{.5\textwidth}
  \centering
  \includegraphics[width=\linewidth]{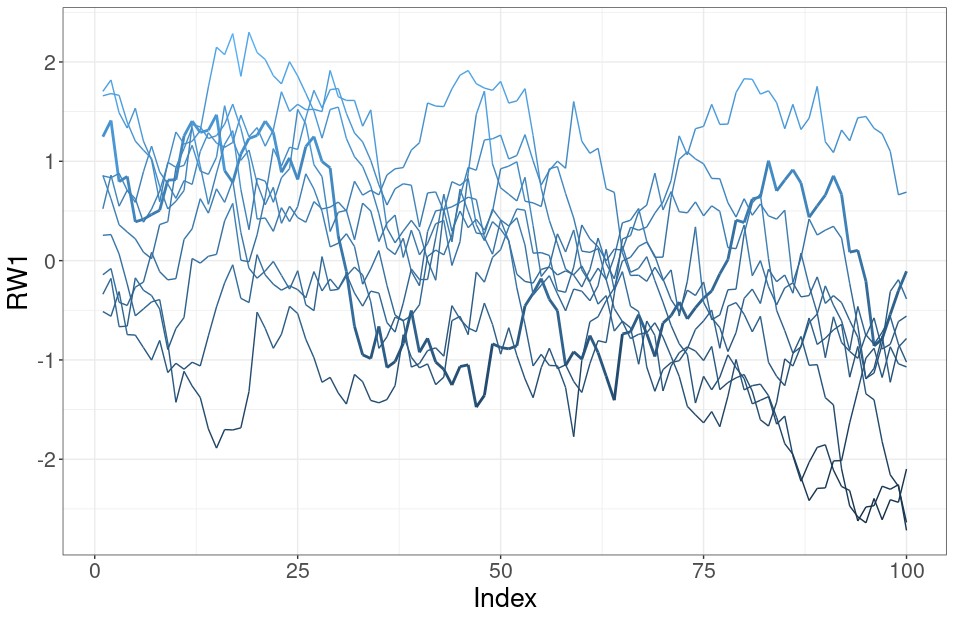}
  %\caption{A subfigure}
  %\label{fig:sub1}
\end{subfigure}%
\begin{subfigure}{.5\textwidth}
  \centering
  \includegraphics[width=\linewidth]{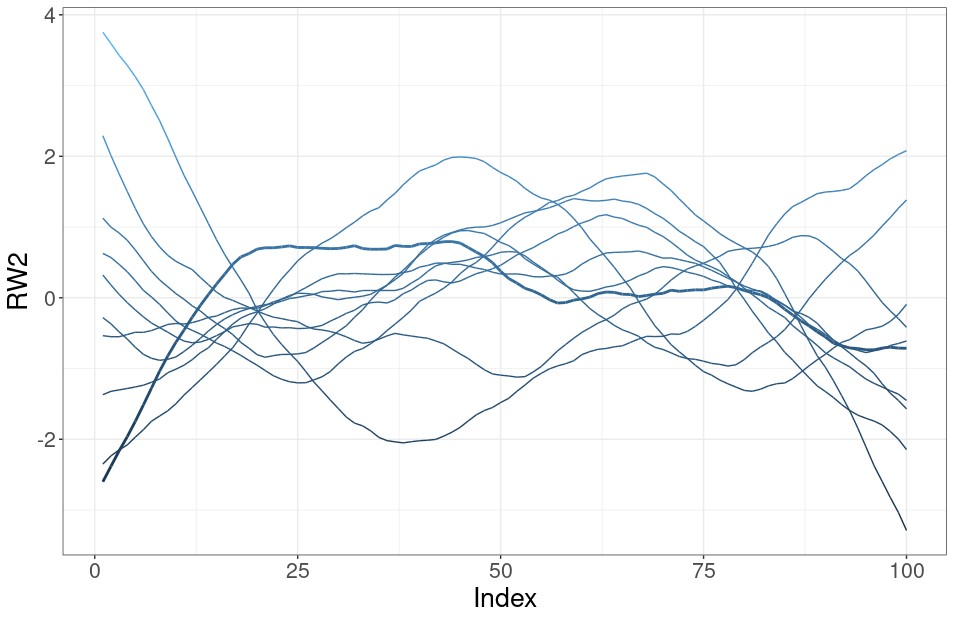}
  %\caption{A subfigure}
  %\label{fig:sub2}
\end{subfigure}
\caption{Illustration of RW1 and RW2 with n =100 for 10 samples each.}
     \label{fig:RW1AND2}
\end{figure}

\noindent RW1 models the difference between two-time points $(\alpha_i - \alpha_{i-1})$ as a process for smooth effects, whereas RW2 models the differences between the two consecutive differences, $(\alpha_{i+2} - \alpha_{i+1}) - (\alpha_{i+1} - \alpha_{i})$, as a process of smoother effects. This dependency makes the precision matrix improper. To obtain a proper model, some constraints are imposed: the sum-to-zero constraint $\sum_i \alpha_i = 0$ for both models and additionally $\sum_i i\alpha_i = 0$ for RW2. Samples from RW2 are smoother than those from RW1, as illustrated in Figure \ref{fig:RW1AND2}.

\vspace{0.3cm}
\noindent \underline{\textbf{Intrinsic Auto-Regressive (ICAR) - Besag Model}}

\noindent The random effect $\pmb \gamma$ is modeled as an ICAR or Besag (Besag, York and Mollié) \cite{Besag1991BayesianIR}. The ICAR model is a type of spatial random effects with a region-based data form, and it is IGMRF,
\begin{equation}
        \pi(\pmb \gamma|\tau_\gamma) \propto \exp\Big(-\dfrac{\tau_\gamma}{2} \displaystyle \sum_{i \sim j} (\gamma_i - \gamma_{j})^2 \Big), 
    \end{equation}
and the conditional distribution,
\begin{equation}
    \gamma_i |\gamma_j , \tau_\gamma \sim \mathcal{N} \Big(\frac{1}{n_i} \sum_{i \sim j} \gamma_j, \frac{1}{n_i \tau_\gamma}\Big)
\end{equation}
where $n_i$ is the number of neighbors of node $i$ and $i \sim j$ represents the neighbors of $i$, $i \neq j$. The mean of node $i$ given its neighbors is the average of the neighbors and the precision is proportional to the neighbors. This model is improper due to the effect of the intercept, thus a sum-to-zero constraint $\sum_i \gamma_i = 0$ is added to make this model identifiable. Another proposed model for the structured spatial effect is the Bym2 model (model = "bym2" in \texttt{R-INLA}), a reparameterization of the Besag-model, which is a sum of the Besag model and an iid model \cite{Simpson2014PenalisingMC}.

\subsection{Interaction Types: \rom{1}, \rom{2}, \rom{3}, and \rom{4}} \label{4intertypes}

\begin{figure}[hbt!]
        \centering
        \includegraphics[width=0.6\textwidth]{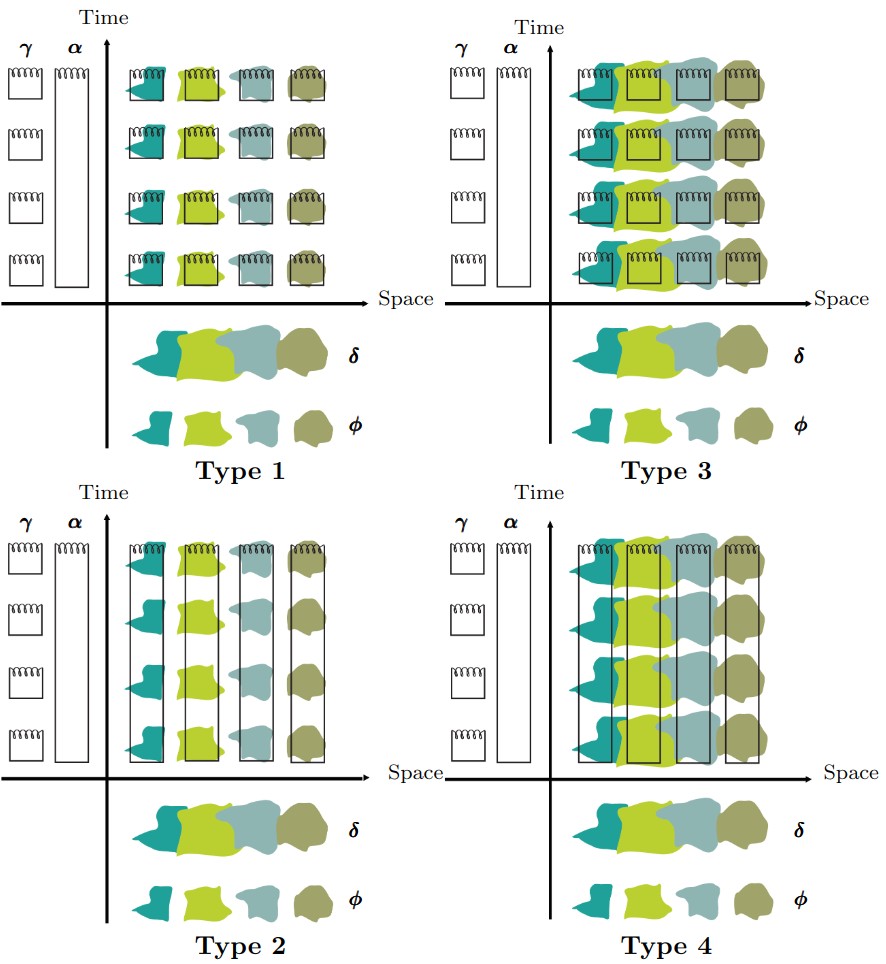}
        \caption{Illustration of the four interaction types}
        \label{fig:illtyp1234}
    \end{figure}

The interaction effect $\pmb \varepsilon$ captures the variation that cannot be explained by the time and space effects. Next, we summarize the four different possibilities for the interaction effect proposed by \cite{KnorrHeld2000BayesianMO}. The idea is to combine one of the two temporal effects with one of the two spatial effects. The four interaction types are illustrated in figure \ref{fig:illtyp1234}. We represent the structured time effect by one calendar and the unstructured time effect by separate calendars. We do the same for space by considering the connected map as a structured spatial effect. Each box in figure \ref{fig:illtyp1234} represents a possible interaction between time and space. The possibilities proceed from incomplete independence to complete dependence. 

\vspace{0.3cm}
\noindent \underline{\textbf{Type \rom{1} Interaction}}

\noindent All interaction parameters $\varepsilon_{ij}$ between the unstructured blocks of $\pmb \gamma$ and $\pmb \phi$ are a priori independent,
\begin{equation}
    \pi(\pmb \varepsilon|\tau_\varepsilon) \propto \exp\Big(-\dfrac{\tau_\varepsilon}{2} \sum_{i= 1}^{n} \sum_{j=1}^{m} \varepsilon_{ij}^2\Big),
\end{equation}
and they present no spatial or temporal correlations. The structure matrix $\pmb Q_\varepsilon = \pmb I_n \otimes \pmb I_m$ has rank $nm$ and the number of constraints on the latent field $\pmb x$ is $l+1$.

\vspace{0.3cm}
\noindent \underline{\textbf{Type \rom{2} Interaction}}

\noindent In this type, we assume there is an interaction between the main effect $\pmb \alpha$ that has a structured block and $\pmb \phi$ that has an unstructured block, such that $\pmb \varepsilon_j = (\varepsilon_{j1},\varepsilon_{j2},\ldots,\varepsilon_{jn})^T$, $j = 1, \ldots m$, follow a random walk independently of all other counties,
\begin{equation}
    \pi(\pmb \varepsilon|\tau_\varepsilon) \propto \exp\Big(  - \dfrac{\tau_\varepsilon}{2} \sum_{j= 1}^{m} \sum_{i=2}^{n} (\varepsilon_{ji} - \varepsilon_{j,i-1})^2 \Big).
\end{equation}
Each area has a temporal correlation structure, but neighboring areas have independent temporal correlations. The structure matrix $\pmb R_\varepsilon = \pmb R_\alpha \otimes \pmb I_m$ has rank $m(n-l)$ and the number of constraints on the latent field $\pmb x$ is $lm+l+1$.
%The temporal effects are different from county to county but they do not have any structure in space.

\vspace{0.3cm}
\noindent \underline{\textbf{Type \rom{3} Interaction}}

\noindent Assume that the main effect $\pmb \delta$, that has structured block, and $\pmb \gamma$, that has unstructured block, interact. Then each $\pmb \varepsilon_i = (\varepsilon_{1i},\varepsilon_{2i},\ldots,\varepsilon_{mi})^T$, $i = 1, \ldots n$ will follow an independent intrinsic auto-regression,
\begin{equation}
    \pi(\pmb \varepsilon|\tau_\varepsilon) \propto \exp\Big(-\dfrac{\tau_\varepsilon}{2} \sum_{i = 1}^{n} \sum_{j \sim k} (\varepsilon_{ji} - \varepsilon_{ki})^2\Big).
\end{equation}
The spatial trends are different from time point to time point, but there is a spatial correlation at each time point. The structure matrix $\pmb R_\varepsilon = \pmb I_m \otimes \pmb R_\delta$ has rank $n(m-1)$ and the number of constraints on the latent field $\pmb x$ is $n+l+1$.

\vspace{0.3cm}
\noindent \underline{\textbf{Type \rom{4} Interaction}}

\noindent Assuming that the two main effects $\pmb \alpha$ and $\pmb \beta$ interact. Then, $\pmb \varepsilon$ can be more factorized into independent blocks,
\begin{equation}
    \pi(\pmb \varepsilon|\tau_\varepsilon) \propto \exp\Big(  - \dfrac{\tau_\varepsilon}{2} \sum_{t = 2}^{T} \sum_{i \sim j} (\varepsilon_{it} - \varepsilon_{jt} - \varepsilon_{it-1} - \varepsilon_{jt-1})^2 \Big).
\end{equation}
At each time point and for contiguous periods, there are spatial correlations, and vice-versa. The structure matrix $\pmb R_\varepsilon = \pmb R_\alpha \otimes \pmb R_\delta$ has a rank $(n-l)(m-1)$ and the number of constraints on the latent field $\pmb x$ is $n + lm - 1$.

\vspace{0.3cm}

The interaction term in the four types can reveal the dependency between space and time, and the spatiotemporal variation of a particular disease. Analyzing the interaction term can contribute to informative decisions during the course of a livestock disease epidemic and improves our understanding of the spread of an infection \cite{Picado2007SpacetimeIA}. Other applications include modeling the variation of the associations between mental illness and substance use mortality and unemployment \cite{Song2022TheSA} and investigating the associations between bone diseases and community water fluoridation \cite{Lee2020TheAB}.

\subsection{Intrinsic Model Scaling} \label{scalingPM}

Having more than IGMRF in the model makes the interpretation of the hyperparameters hard if the full model is not scaled. A chosen hyperparameter for a local precision parameter scales one IGMRF, and its value regulates the degree of smoothness of the resulting latent field. Its influence on the local IGMRF depends on its type and the size of the structure matrix. When the value of the precision is small, the field becomes less smooth and the relationships between the neighbors become weaker. \cite{Srbye2014ScalingIG} suggests a way to account for the differences of the marginals of these assigned hyperpriors by scaling the marginal variances to have a unit geometric mean. Assume the graph structure $\pmb R$ of the spatial effect is fully connected, the precision matrix $\pmb Q = \tau_\theta \pmb R$ is scaled as proposed by \cite{FreniSterrantino2018ANO} using
\begin{equation}
    \exp \Big(\dfrac{1}{n} \sum_i \log \text{diag}(R_{i,i}^+) \Big),
\end{equation}
where $\pmb R^+$ is the pseudo inverse of $\pmb R$. After scaling, the marginal variance of each element in the latent field is on average equal to one. When an IGMRF is constructed by a Kronecker product of two structure matrices, it is enough to scale these two matrices before computing the product. \cite{FreniSterrantino2018ANO} also discusses how to scale a graph that is not fully connected. 

\subsection{Prior Distributions for the Hyperparameter}
The hyperparameter $\pmb \theta = (\log \tau_\alpha$, $\log \tau_\gamma$, $\log \tau_\delta$, $\log \tau_\phi$, $\log \tau_\varepsilon$) represents the variations of the blocks in the precision matrix. These parameters are estimated from the data and the commonly used priors are a gamma distribution and PC prior for $\tau$ \cite{Simpson2014PenalisingMC}. The gamma distribution causes overfitting, which can be avoided by the use of PC priors \cite{Simpson2014PenalisingMC}. Different choices of priors are also discussed in the literature \cite{Simpson2014PenalisingMC,Gelman2004PriorDF, Wakefield2007DiseaseMA}. 

\cite{FrancoVilloria2022VariancePI} proposes a new parameterization of the priors for the interaction types \rom{1}, \rom{2}, \rom{3}, and \rom{4} called variance partitioning. This parameterization includes a mixing parameter that balances the main and interaction effects, which improves interpretability. The hyperparameter $\pmb \theta $ can have a joint prior distribution as the main effects and the interactions are not independent. This parameterization allows quantification of the contribution of the total variance into the main effects and the interactions.

\section{New Approach for Fitting Models with Interaction Types \rom{1}, \rom{2}, \rom{3}, and \rom{4}} \label{newapproach}

The new approach uses the same methodology in INLA to approximate Bayesian inference. However, it does not have the sparsity assumption for matrices, and it corrects for the constraints using the pseudo-inverse of the precision matrix. We outline in Appendix \ref{Appendix A} the main key steps used in INLA approach to approximate inference for intrinsic models, and we summarize its two sequential stages. 

The latent field $\pmb x$ in the new approach is composed of the linear predictor $\pmb x = \pmb \eta$ in the first stage and the effects $\pmb x = \pmb e$ in the second stage. This gives the computational advantage in the first stage for the models where the size of the data is less than the size of the effects, and it is the case when fitting disease-mapping models. We present in this section the approach used in the new approach to update the covariance matrix under linear constraints. This approach is used to compute the marginal posteriors of the effects and linear predictor.

\subsection{IGMRF under Sum-to-Zero Constraints}

Assume $\pmb x$ is an IGMRF of order $k$, mean $\pmb \mu = \pmb 0$ and covariance matrix $\pmb \Sigma \in \mathbb{R}^s$. Let $\lambda_1, \ldots, \lambda_{s}$ be the eigenvalues, and $\pmb v_1, \ldots, \pmb v_{s}$ be the corresponding eigenvectors of the eigen-decomposition of $\pmb \Sigma$. The log-density of this Gaussian field conditioned on set of constraints $\pmb C \in \mathbb{R}^{k \times s}$ such that $\pmb C \pmb x = \pmb 0$,
\begin{equation}
    \log \pi (\pmb x | \pmb C \pmb x = \pmb 0) = \frac{s - k}{2}\log 2\pi + \dfrac{1}{2} \sum_{i=1}^{s-k} \log \lambda_i - \dfrac{1}{2} \pmb x^T \pmb {\Tilde{\Sigma}}^{+} \pmb x,
    \label{conditionConstraints}
\end{equation}
where $\pmb {\Tilde{\Sigma}}^{+}$ is the pseudo-inverse of the covariance matrix of the latent field $\pmb x$ conditioned on $\pmb C \pmb x$, and we denote $\pmb {\Tilde{\Sigma}}$ by $\text{cov}(\pmb x | \pmb C \pmb x)$. Using kriging technique, the conditioned covariance $\pmb {\Tilde{\Sigma}}$ is,
\begin{equation}
    \pmb {\Tilde{\Sigma}} = \pmb \Sigma - \pmb \Sigma \pmb C^T (\pmb C \pmb \Sigma \pmb C^T)^{-1} \pmb C \pmb \Sigma.
\end{equation}
The set of imposed constraints spans the null space of $\pmb {\Tilde{\Sigma}}$,
\begin{equation}
    \Big( \pmb \Sigma - \pmb \Sigma \pmb C^T (\pmb C \pmb \Sigma \pmb C^T)^{-1} \pmb C \pmb \Sigma \Big) \pmb C^T = \pmb 0,
\end{equation}
which makes $\pmb {\Tilde{\Sigma}}$ equivalent to,
\begin{equation}
    \pmb {\Tilde{\Sigma}} = \sum_i^{s-k} \lambda_i \pmb v_i \pmb v_i^T.
\end{equation}
$\pmb {\Tilde{\Sigma}}$ is also defined as the generalized inverse of the improper precision matrix of $\pmb x$, and we denote it by $\pmb Q^+$ for the rest of this paper. The use of the dense matrix $\pmb Q^+$ corrects for the imposed constraints $\pmb C \pmb x = \pmb 0$ such that $\pmb C$ is composed of the eigenvectors $\{\pmb v_i\}_{s-k+1}^s$ that correspond to zero eigenvalues when decomposing the precision matrix of $\pmb Q$ or infinite eigenvalues when decomposing the covariance matrix $\pmb \Sigma$.

\subsection{Full Conditional Distribution of an IGMRF under Linear Constraints}

\noindent The full conditional distribution of the IGMRF $\pmb x$ under linear constraints $\pmb C \pmb x = \pmb 0$, given $\pmb \theta$, is
\begin{equation}
     \pi(\pmb x|\pmb \theta,\pmb y, \pmb C \pmb x = \pmb 0) = \pi(\pmb y|\pmb x,\pmb \theta) \pi(\pmb x|\pmb \theta, \pmb C \pmb x = \pmb 0).
\end{equation}
Usually $\pi(\pmb y|\pmb x,\pmb \theta)$ is not Gaussian and $\pi(\pmb x|\pmb \theta, \pmb C \pmb x = \pmb 0)$ is a Gaussian prior distribution of the latent field with zero mean and generalized inverse $\pmb {\tilde{\Sigma}_x}$ of its precision matrix. To approximate this marginal posterior by Gaussian distribution of covariance matrix $\pmb \Sigma^{*}$, we write it in this form,
\begin{equation} 
\begin{split}
\tilde{\pi}_{G}(\pmb x|\pmb \theta,\pmb y,\pmb C \pmb x = \pmb 0) & \propto \exp\Big(-\displaystyle\frac{1}{2}(\pmb x - \pmb x^*)^{T} \pmb {\Sigma}^{*+}(\pmb x - \pmb x^*)\Big)
\end{split}
\end{equation}
and after some expansion,
\begin{equation} 
\begin{split}
\tilde{\pi}_{G}(\pmb x|\pmb \theta, \pmb y,\pmb C \pmb x = \pmb 0) & \propto \exp\Big(-\displaystyle\frac{1}{2}\pmb x^{T} \pmb \Sigma^{*+}\pmb x + \pmb x^{*T} \pmb \Sigma^{*+} \pmb x\Big) = \exp\Big(-\displaystyle\frac{1}{2}\pmb x^{T} \pmb \Sigma^{*+} \pmb x + \pmb b^T \pmb x \Big)
\end{split}
\end{equation}
where,

\begin{equation}
    \pmb b = {\pmb x^*}^T \pmb \Sigma^{*+}  = \nabla g(\pmb x) - \nabla^2 g(\pmb x) \pmb \mu_{l},
    \label{GAequation}
\end{equation}

and $\pmb \mu_{l}$ is what maximizes the likelihood and since the latent field $\pmb x$ is maximized at zero, then ${\pmb x^*}$ also maximizes the full conditional latent field and it becomes,

\begin{equation}
    \pmb b = \nabla g (\pmb x) - \nabla^2 g(\pmb x) \pmb x^*.
\end{equation}

The covariance $\pmb \Sigma^{*}$ is updated using Woodbury formula \cite{Riedel1992ASI},

\begin{equation}
    \pmb \Sigma^{*} = \pmb Q_{x}^+ - \pmb Q_{x}^+(\pmb I +  \pmb Q_{l}\pmb Q_{x}^+)^{-1}\pmb Q_{l} \pmb Q_{x}^+,
\end{equation}
%\begin{equation}
%    \pmb \Sigma^{*} = \Big(\pmb Q_l + {\tilde{\pmb \Sigma}_x}^+\Big)^+,
%\end{equation}

where $\pmb Q_{l}$ is the second-order partial derivatives of the likelihood $\pi(\pmb y|\pmb x,\pmb \theta)$ with respect to $\pmb x$. Equation (\ref{GAequation}) is solved after some iterations before getting the Gaussian approximation,

\begin{equation}
    \pmb x|\pmb \theta,\pmb y,\pmb C \pmb x = \pmb 0 \sim \mathcal{N}(\pmb x^*,\pmb {\Sigma}^*).
\end{equation}

\textbf{Example:} We compute here the precision matrix $\pmb {\Sigma}^*$ in two ways: kriging technique and our proposed approach. Given the Poisson model,
\begin{equation}
    \pmb y \sim \text{Poisson}(\pmb \eta = \pmb A \pmb e),
\end{equation}
where $\pmb y$ is the observed count points, $\pmb \eta$ is the linear predictor, $\pmb e$ are the effects and $\pmb A \in \mathbb{R}^{4x3}$ is the mapping matrix, $\{( (1, 1, 0, 0), (1, 0, 1, 0), (0, 0, 0, 1) \}$. We assume the precision matrices,

\begin{equation}
\pmb Q_{l}(\pmb \eta) = - \dfrac{\partial^2 \pi(\pmb y| \pmb \eta^2)}{\partial \pmb \eta^2} = \text{diag}(1.796,2.033,0.896)
\text{ and }  
\pmb Q_{\pmb e} = \begin{pmatrix}
~~~1 & -1 & ~~~0 & ~~~0\\
-1 & ~~~2 & -1 & ~~~0\\
~~~0 & -1 & ~~~2 & -1\\
~~~0 & ~~~0 & -1 & ~~~1
\end{pmatrix}.
\end{equation}

The imposed constraint is $\pmb C \pmb e = (1~~1~~1~~1) ~ \pmb e  = 0$. We compute the uncorrected precision matrix,

\begin{equation}
    \pmb \Sigma^*_{un} = (\pmb A^T \pmb Q_l \pmb A + \pmb Q_{\pmb e} + \varepsilon \pmb I)^{-1} = \begin{pmatrix}
~~~0.350 & -0.150 & -0.293 & -0.320\\
-0.150 & ~~~0.350 & ~~~0.207 & ~~~0.180\\
-0.293 & ~~~0.207 & ~~~0.554 & ~~~0.430\\
-0.320 & ~~~0.180 & ~~~0.430 & ~~~0.905
\end{pmatrix},
\end{equation}

where $\varepsilon$ is a tiny noise (say $1e^{-4}$). Using equation (\ref{kriggCov}), we correct for the constraints to get, 

\begin{equation}
    \pmb \Sigma^* = \begin{pmatrix}  ~~~0.274 & -0.044 &-0.129 &-0.102 \\
 -0.044 &~~~0.198 &-0.025 &-0.129 \\
 -0.129 &-0.025 &~~~0.198 &-0.044 \\
 -0.102 &-0.129 &-0.044 &~~~0.274 \end{pmatrix}.
\end{equation}
Equivalently, we get the same results using the Woodbury formula, 

\begin{equation}
  \pmb \Sigma^* = \pmb Q^{+}_{\pmb e} - \pmb Q^{+}_{\pmb e} (\pmb I + \pmb A^T \pmb Q_l \pmb A \pmb Q^{+}_{\pmb e})^{-1} \pmb A^T \pmb Q_l \pmb A \pmb Q^{+},
\end{equation}

where $\pmb Q^{+}_{\pmb e}$ is the pseudo-inverse of $\pmb Q_{\pmb e}$,
\begin{equation}
    \pmb Q^{+}_{\pmb e} = \begin{pmatrix}  ~~~0.875 & ~~~0.125 & -0.375 & -0.625\\
~~~0.125 & ~~~0.375 & -0.125 & -0.375\\
-0.375 & -0.125 & ~~~0.375 & ~~~0.125\\
-0.625 & -0.375 & ~~~0.125 & ~~~0.875 \end{pmatrix}.
\end{equation}

\subsection{Equivalence to Sum-to-Zero Constraints}
Imposing constraints to the models is not an easy task. We revisit again the interaction types \rom{1}, \rom{2}, \rom{3} and \rom{4} using the notation in equation (\ref{formblockdiagonal}), and we assume the following cases:

\textbf{Case 1:} $\pmb R_\alpha$ is a structure matrix of RW2, and $\pmb R_\delta$ is a structure matrix of one full connected graph, every county or area has at least one neighbor.\\
\textbf{Case 2:} $\pmb R_\alpha$ is a structure matrix of RW2, and $\pmb R_\delta$ is a structure matrix of an unconnected graph that is divided into connected three sub-graph of sizes $p_1, p_2$, and $p_3$, such that $\pmb \delta = (\delta_1, \ldots, \delta_{p_1}, \delta_{p_1 + 1}, \ldots, \delta_{p_1 + p_2}, \delta_{p_1 + p_2 + 1}, \ldots, \delta_{m})$. For example three islands, the first island has $p_1$ connected areas, the second island has $p_2$ connected areas, and the third has $p_3$ connected areas.

Consider interaction type \rom{1}, the imposed constraints in case 1,
\begin{equation}
    \sum_i^n \alpha_i = 0, \quad \sum_i^n i\alpha_i = 0, \quad \sum_{i=1}^{m} \delta_{i} = 0, 
\end{equation}
are equivalent to using the structures $\pmb R_\alpha^{+}$ and $\pmb R_\delta^{+}$ in the computations. Similarly, in case 2, the constraints
\begin{equation}
    \sum_i^n \alpha_i = 0, \quad \sum_i^n i\alpha_i = 0, \quad \sum_{i=1}^{p_1} \delta_{i} =0, \quad \sum_{i=p_1 + 1}^{p_1 + p_2} \delta_{i} = 0, \quad \text{and} \quad \sum_{i=p_1 + p_2 + 1}^{m} \delta_{i} = 0\\
\end{equation}
are equivalent to using the structures $\pmb R_\alpha^{+}$ and $\pmb R_\delta^{+}$ in the computations. Now consider the complex interaction type \rom{4}, the constraints in case 1,
\begin{equation}
\quad \sum_i^n \alpha_i = 0, \quad \sum_i^n i\alpha_i = 0, \quad \sum_{i=1}^{m} \delta_{i} = 0,
\end{equation}
\begin{equation}
\quad \sum_{j=1}^{n}  \varepsilon_{i,j} = 0, i =1, \ldots, m, \quad \sum_{i=1}^{m} \varepsilon_{i,j} = 0, j =1, \ldots, n -1, \quad \text{and} \quad \sum_{j=1}^{n}  j \varepsilon_{i,j} = 0, i =1, \ldots, m - 1
\end{equation}
are equivalent to using the structures $\pmb R_\alpha^{+}$, $\pmb R_\delta^{+}$ and $\pmb R_\varepsilon^{+}$ in the computations. These three structures are used also in case 2, and similar to previous cases we can find the equivalence sum-to-zero constraints. If constraints are not well specified, results and the interpretation of parameters will be wrong. The new approach can be used easily as an inferential tool for a variety of spatiotemporal applications with the interactions types \rom{1}, \rom{2}, \rom{3} and \rom{4}.

\subsection{A Hybrid Parallel Computing Approach}

We exploits the architecture of multi-core systems and maximizes the performance of the approach through a parallel computing implementation using OpenMP and MPI. OpenMP is an Application Program Interface (API) that provides a portable, scalable model for developers of shared-memory parallel applications on most processor architectures and operating systems. MPI is a Message Passing Interface that allows running several processes in parallel on shared-distributed memory, and each process runs with a number of threads. 

\begin{figure}[H]
     \centering
     \begin{subfigure}[b]{0.33\textwidth}
         \centering
         \includegraphics[width=\textwidth]{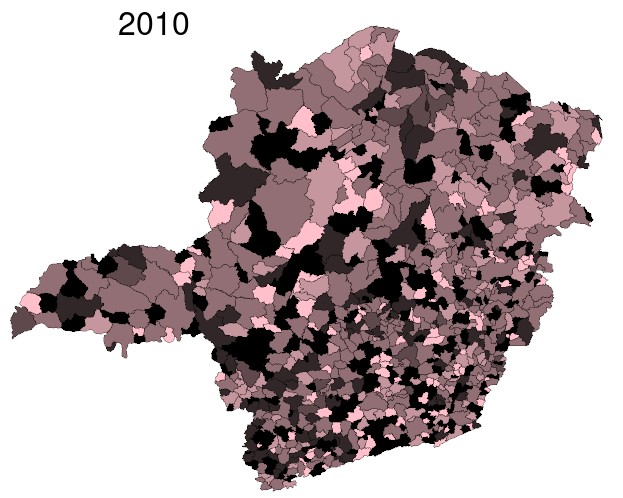}
     \end{subfigure}
     \hfill
     \begin{subfigure}[b]{0.33\textwidth}
         \centering
         \includegraphics[width=\textwidth]{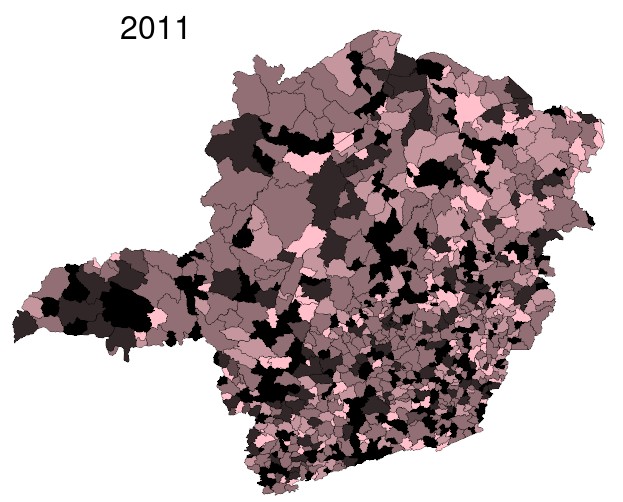}
     \end{subfigure}
     \hfill
     \begin{subfigure}[b]{0.33\textwidth}
         \centering
         \includegraphics[width=\textwidth]{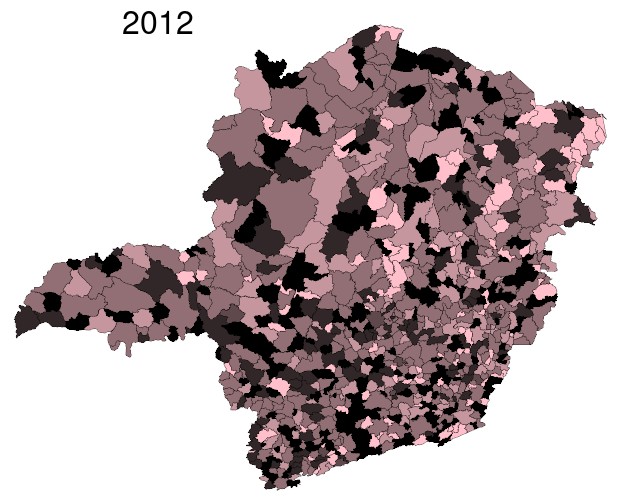}
     \end{subfigure}
     \qquad
    \begin{subfigure}[b]{0.33\textwidth}
         \centering
         \includegraphics[width=\textwidth]{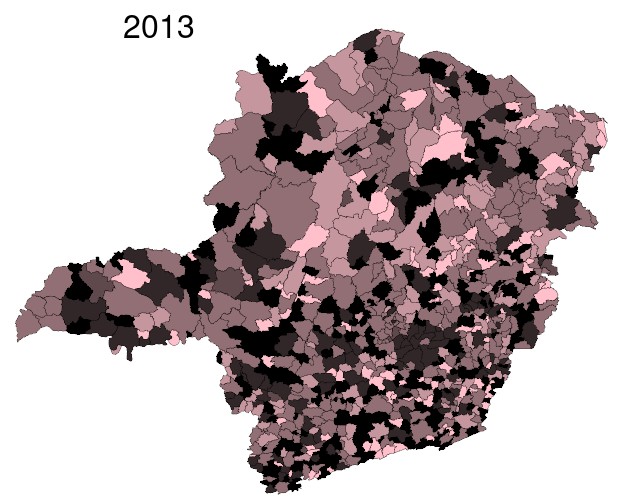}
     \end{subfigure}
     \hfill
     \begin{subfigure}[b]{0.33\textwidth}
         \centering
         \includegraphics[width=\textwidth]{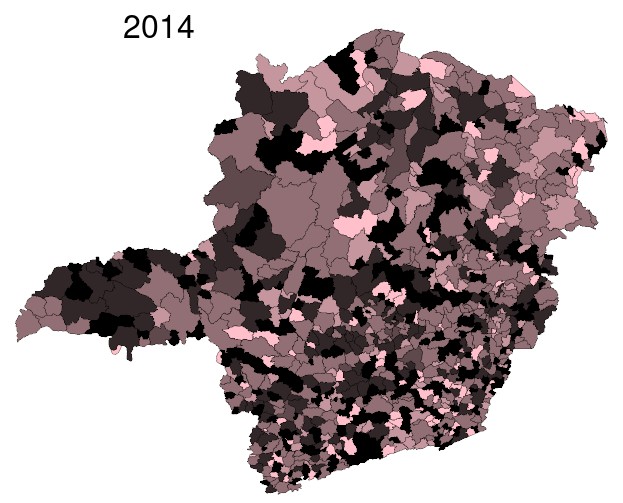}
     \end{subfigure}
     \hfill
     \begin{subfigure}[b]{0.3\textwidth}
         \centering
         \includegraphics[width=\textwidth]{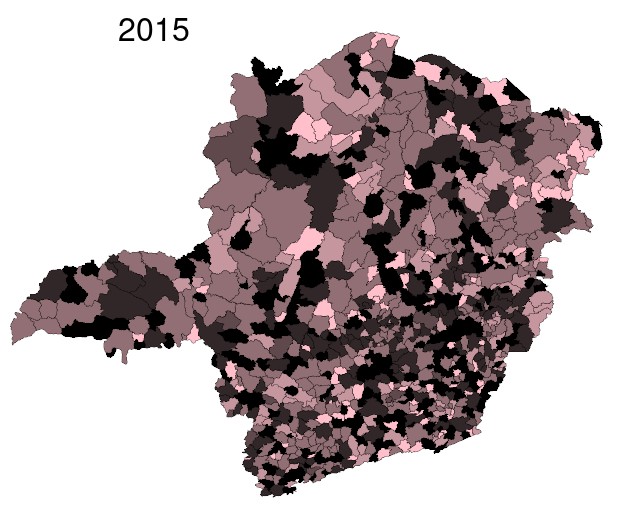}
     \end{subfigure}
     \qquad
     \begin{subfigure}[b]{0.33\textwidth}
         \centering
         \includegraphics[width=\textwidth]{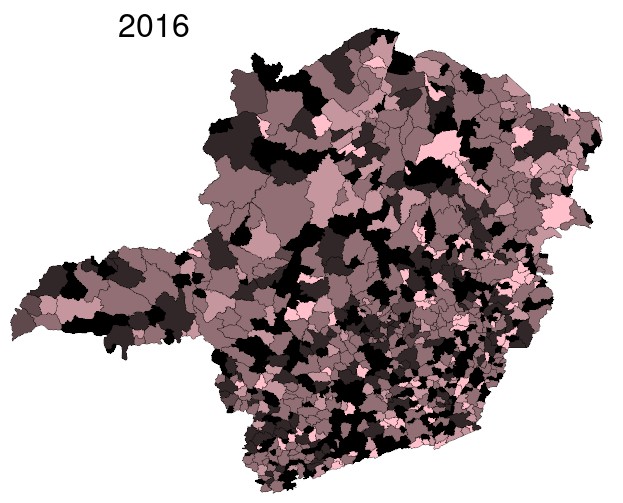}
     \end{subfigure}
     \hfill
     \begin{subfigure}[b]{0.33\textwidth}
         \centering
         \includegraphics[width=\textwidth]{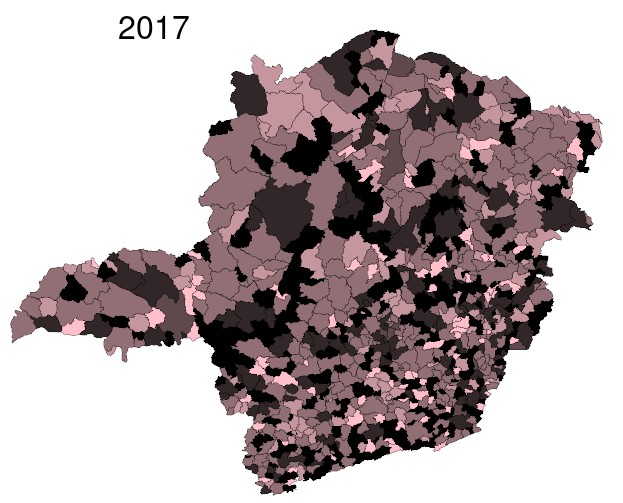}
     \end{subfigure}
     \hfill
     \begin{subfigure}[b]{0.3\textwidth}
         \centering
         \includegraphics[width=\textwidth]{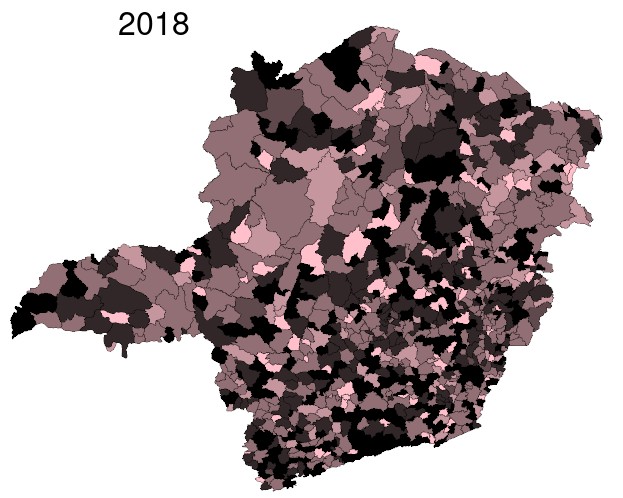}
     \end{subfigure}
     \qquad
     \begin{subfigure}[b]{0.33\textwidth}
         \centering
         \includegraphics[width=\textwidth]{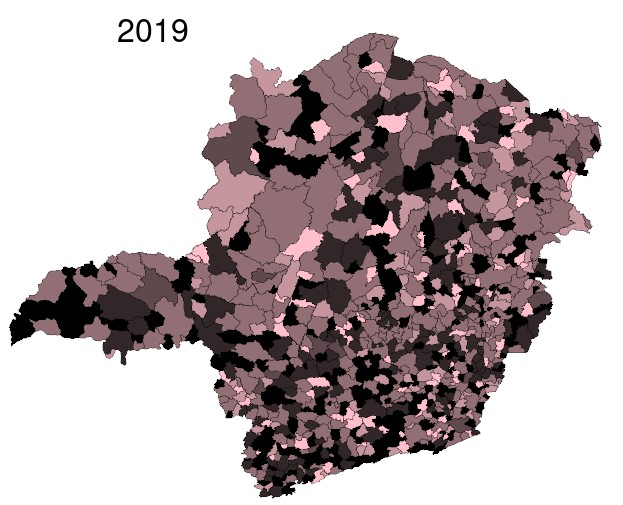}
     \end{subfigure}
     \hfill
     \begin{subfigure}[b]{0.33\textwidth}
         \centering
         \includegraphics[width=\textwidth]{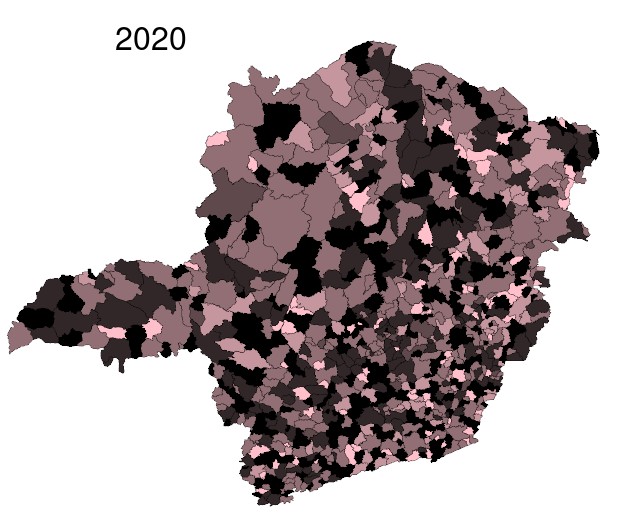}
     \end{subfigure}
     \hfill
     \begin{subfigure}[b]{0.33\textwidth}
         \centering
         \includegraphics[scale = 0.2]{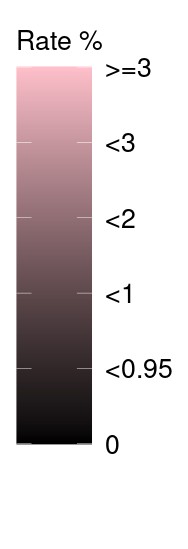}
     \end{subfigure}
        \caption{Infant Mortality Rate $\%$ in Minas Gerais State from 2010 to 2020}
        \label{fig:maps}
\end{figure}

The new approach is written in C\small{++} \normalsize{} using Blaze library, an open-source, high-performance math library for dense linear algebra \cite{blaze2}. Its careful implementation brings out a dense version of integrated nested Laplace approximations, and combines an intuitive user interface with high-performance speed. The new approach is scaled up to the explored points of the hyperparameter $\pmb \theta$. For instance, given $n_1$ explored points, $n_2$ nodes, and $n_3$ cores per node, we can run $n_1$ processes on the available nodes, and each process is given $\lfloor n_2/n_3 \rfloor$ or $\lfloor (n_2n_3)/h \rfloor$ cores, such that $h$ is the dimension of the hyperparameter.

\section{Applications}
\subsection{Simulation Study}

% \usepackage{multirow}

In this section, we compare computationally the two methodologies INLA and the new approach by fitting Poisson observations with considerable number of interactions. We fit a generalized additive model with a linear predictor,
\begin{equation}
    \eta_{ij} = \mu  + \alpha_i + \gamma_i + \delta_j + \phi_j + \varepsilon_{ij}, ~~~i = 1, \ldots n, j = 1, \ldots m,
    \label{sim1_example}
\end{equation}
where $\mu$ is the overall intercept with precision 0.001, $\pmb \alpha$ is a random walk of order 2 of size $n$, and $\pmb \gamma$ is the Besag model of size $m$, $\gamma$ and $\phi$ are independent and identically distributed Gaussian with precision matrices 250 $\pmb I_n$ and 250 $\pmb I_m$ respectively, and $\pmb \varepsilon$ is the interaction term of type \rom{4}. The precision matrix is scaled and a PC-prior is used for the hyperparameters. The number of the needed constraints to get precise inference is $k = nm - (n-2)(m-1) + 3$, see Section \ref{4intertypes}. We simulate the observed values using the precision matrices of the effects, and create a one-connected graph for the Besag model such that the sum of rows and columns sum-to-zero, see \cite{Rue2005GaussianMR} for algorithms.
{
\renewcommand{\arraystretch}{1.4}
\begin{table}[H]
\centering
\resizebox{\columnwidth}{!}{%
\begin{tabular}{|c|c|c|c|c|ccc|ccc|}
\hline
\multirow{3}{*}{\textbf{\begin{tabular}[c]{@{}c@{}}Space\\ size\end{tabular}}} & \multirow{3}{*}{\textbf{\begin{tabular}[c]{@{}c@{}}Time\\ size\end{tabular}}} & \multicolumn{1}{l|}{\multirow{3}{*}{\textbf{\begin{tabular}[c]{@{}l@{}}Effects\\ size (s)\end{tabular}}}} & \multirow{3}{*}{\textbf{\begin{tabular}[c]{@{}c@{}}Constraints\\ (k)\end{tabular}}} & \multirow{3}{*}{\textbf{\begin{tabular}[c]{@{}c@{}}Ratio\\ (s/k)\end{tabular}}} & \multicolumn{3}{c|}{\textbf{Gaussian Approximation}}                                                                                                                                                                                                                                        & \multicolumn{3}{c|}{\textbf{Variational Bayes Correction}}                                                                                                                                                                                                                                       \\ \cline{6-11} 
                                                                               &                                                                               & \multicolumn{1}{l|}{}                                                                                     &                                                                                     &                                                                                 & \multicolumn{1}{c|}{\textbf{\begin{tabular}[c]{@{}c@{}}\texttt{inla()}\\ (1 node)\end{tabular}}} & \multicolumn{1}{c|}{\textbf{\begin{tabular}[c]{@{}c@{}}\texttt{inla1234()}\\ (25 nodes)\end{tabular}}} & \multirow{2}{*}{\textbf{\begin{tabular}[c]{@{}c@{}}Speedup\\ Ratio\end{tabular}}} & \multicolumn{1}{c|}{\textbf{\begin{tabular}[c]{@{}c@{}}\texttt{inla()}\\ (1 node)\end{tabular}}} & \multicolumn{1}{c|}{\textbf{\begin{tabular}[c]{@{}c@{}}\texttt{inla1234()}\\ (25 nodes)\end{tabular}}} & \multirow{2}{*}{\textbf{\begin{tabular}[c]{@{}c@{}}Speedup\\ Ratio\end{tabular}}} \\ \cline{6-7} \cline{9-10}
                                                                               &                                                                               & \multicolumn{1}{l|}{}                                                                                     &                                                                                     &                                                                                 & \multicolumn{2}{c|}{\textbf{Execution Time (s)}}                                                                                                                                    &                                                                                   & \multicolumn{2}{c|}{\textbf{Execution Time (s)}}                                                                                                                                    &                                                                                   \\ \hline
100                                                                            & 10                                                                            & 1221                                                                                                      & 211                                                                                 & 5.78                                                                            & \multicolumn{1}{c|}{37.68}                                                            & \multicolumn{1}{c|}{2.89}                                                                   & 13.03                                                                             & \multicolumn{1}{c|}{56.06}                                                            & \multicolumn{1}{c|}{3.23}                                                                   & 17.35                                                                             \\ \hline
100                                                                            & 20                                                                            & 2241                                                                                                      & 221                                                                                 & 10.14                                                                           & \multicolumn{1}{c|}{57.18}                                                            & \multicolumn{1}{c|}{17.12}                                                                  & 3.33                                                                              & \multicolumn{1}{c|}{241.39}                                                           & \multicolumn{1}{c|}{19.36}                                                                  & 12.46                                                                             \\ \hline
100                                                                            & 30                                                                            & 3261                                                                                                      & 231                                                                                 & 14.11                                                                           & \multicolumn{1}{c|}{84.08}                                                            & \multicolumn{1}{c|}{41.08}                                                                  & 2.04                                                                              & \multicolumn{1}{c|}{703.94}                                                           & \multicolumn{1}{c|}{68.53}                                                                  & 10.27                                                                             \\ \hline
100                                                                            & 40                                                                            & 4281                                                                                                      & 241                                                                                 & 17.7                                                                            & \multicolumn{1}{c|}{160.21}                                                           & \multicolumn{1}{c|}{75.42}                                                                  & 2.12                                                                              & \multicolumn{1}{c|}{1643.13}                                                          & \multicolumn{1}{c|}{107.83}                                                                 & 15.23                                                                             \\ \hline
200                                                                            & 5                                                                             & 1411                                                                                                      & 400                                                                                 & 3.47                                                                            & \multicolumn{1}{c|}{53.06}                                                            & \multicolumn{1}{c|}{2.35}                                                                   & 22.58                                                                             & \multicolumn{1}{c|}{85.76}                                                            & \multicolumn{1}{c|}{3.10}                                                                   & 27.66                                                                             \\ \hline
200                                                                            & 30                                                                            & 6461                                                                                                      & 431                                                                                 & 14.99                                                                           & \multicolumn{1}{c|}{346.83}                                                           & \multicolumn{1}{c|}{285.67}                                                                 & 1.21                                                                              & \multicolumn{1}{c|}{5274.71}                                                          & \multicolumn{1}{c|}{394.18}                                                                 & 13.38                                                                             \\ \hline
400                                                                            & 5                                                                             & 2811                                                                                                      & 806                                                                                 & 3.48                                                                            & \multicolumn{1}{c|}{366.89}                                                           & \multicolumn{1}{c|}{14.06}                                                                  & 26.09                                                                             & \multicolumn{1}{c|}{632.13}                                                           & \multicolumn{1}{c|}{17.94}                                                                  & 35.23                                                                             \\ \hline
400                                                                            & 10                                                                            & 4821                                                                                                      & 811                                                                                 & 5.94                                                                            & \multicolumn{1}{c|}{736.14}                                                           & \multicolumn{1}{c|}{60.12}                                                                  & 12.24                                                                             & \multicolumn{1}{c|}{2540.92}                                                          & \multicolumn{1}{c|}{80.61}                                                                  & 31.52                                                                             \\ \hline
800                                                                            & 5                                                                             & 5611                                                                                                      & 1606                                                                                & 3.49                                                                            & \multicolumn{1}{c|}{1822.32}                                                          & \multicolumn{1}{c|}{82.92}                                                                  & 21.98                                                                             & \multicolumn{1}{c|}{4093.90}                                                          & \multicolumn{1}{c|}{102.33}                                                                 & 40.01                                                                             \\ \hline
\end{tabular}}
\caption{Comparison between INLA and the new approach.}
\label{CINLAINLA1234}
\end{table}
}{ \renewcommand{\arraystretch}{0.5}}

The goal is to compare the execution time needed to obtain inference using INLA and the new approach. Both approaches find the mode of the hyperparameter and estimate the marginal posteriors of the hyperparameter and the latent field using Gaussian approximation and Variational Bayes Correction \cite{Niekerk2021CorrectingTL}.

Table \ref{CINLAINLA1234} shows that the new approach has a significantly faster execution time for computing the marginal posteriors in these types of models. We run INLA on a single Cascade Lake CPU node, 40 cores, 2.50 GHz, 384 GB/usable 350 GB, and we exploit fully the number of threads present (10:4). However, we run the new approach on 25 nodes Cascade Lake nodes. The availability of more nodes fosters a reduction in the execution time, see Table \ref{Morenodes}.

The computations in the new approach depend on the size of the linear predictor, but are independent of the number of constraints. However, INLA depends on the sparsity of the precision matrices, the size of the effects, and the number of constraints. In the presence of adequate computational resources, the presented approach outperforms INLA in the presence of high interactions. The increase in speed ranges from 1.21 to 26.09 when using Gaussian approximation (GA) and 10.27 to 40.01 when using high-order Variational Bayes Correction (VBC), which scales better than Laplace approximation \cite{Niekerk2021CorrectingTL}. The speed depends mainly on the ratio (s/k) and the type of approximation the practitioner wants to achieve by using GA or VBC.

%{ 
%\renewcommand{\arraystretch}{1.4}
%\begin{table}
%\centering
%\resizebox{\columnwidth}{!}{%
%}
%\caption{Comparison between INLA and INLA1234$ approaches. GA: Gaussian Approximation, VBC: %Variational Bayes Correction}
%\label{CINLAINLA1234}
%\end{table}
%}{ \renewcommand{\arraystretch}{1.0}}

{ \renewcommand{\arraystretch}{1.4}
\begin{table} %hbt!
\centering
\begin{tabular}{|c|cc|cc|}
\hline
\multicolumn{1}{|l|}{\multirow{2}{*}{\textbf{Nodes}}} & \multicolumn{2}{c|}{\textbf{Gaussian Approximation}}                & \multicolumn{2}{c|}{\textbf{Variational Bayes Correction}}          \\ \cline{2-5} 
\multicolumn{1}{|l|}{}                                & \multicolumn{1}{c|}{\textbf{Time Execution (s)}} & \textbf{Speedup} & \multicolumn{1}{c|}{\textbf{Time Execution (s)}} & \textbf{Speedup} \\ \hline
1                                                     & \multicolumn{1}{c|}{828.88}                      & 2.20             & \multicolumn{1}{c|}{1335.24}                           & 3.06               \\ \hline
3                                                     & \multicolumn{1}{c|}{231.79}                      & 7.86             & \multicolumn{1}{c|}{317.98}                           & 12.87               \\ \hline
6                                                     & \multicolumn{1}{c|}{190.87}                      & 9.55            & \multicolumn{1}{c|}{268.02}                           & 15.27                \\ \hline
12                                                    & \multicolumn{1}{c|}{180.50}                      & 10.10            & \multicolumn{1}{c|}{232.56}                           & 17.60               \\ \hline
25                                                    & \multicolumn{1}{c|}{82.92}                       & 21.98            & \multicolumn{1}{c|}{102.33}                           & 40.01               \\ \hline
\end{tabular}
\caption{The time execution and speedup ratio when using the new approach (compared to INLA) with different number of nodes for the case: space size = 800, time size = 5, Gaussian approximation: 1822.32 seconds and Variational Bayes Correction: 4093.90 seconds }
\label{Morenodes}
\end{table}
}{ \renewcommand{\arraystretch}{0.5}}

\subsection{Spatiotemporal Variation of Infant Mortality in Minas Gerais State}

We model the spatiotemporal variations of infant mortality in Minas Gerais in Brazil from 2010 to 2020. The infant mortality rate is expressed as the number of deaths per thousand live births. The data can be obtained online from \href{https://datasus.saude.gov.br/informacoes-de-saude-tabnet/}{Informatics Department SUS}. The southeast region of Brazil has the highest contribution to the overall gross domestic product of Brazil. Minas Gerais State is the third highest contributor to gross domestic product (after Sao Paulo and Rio de Janeiro) and is composed of 853 municipalities. 

\begin{figure}[hbt!]
\centering
\begin{subfigure}{.5\textwidth}
  \centering
  \includegraphics[width=\linewidth]{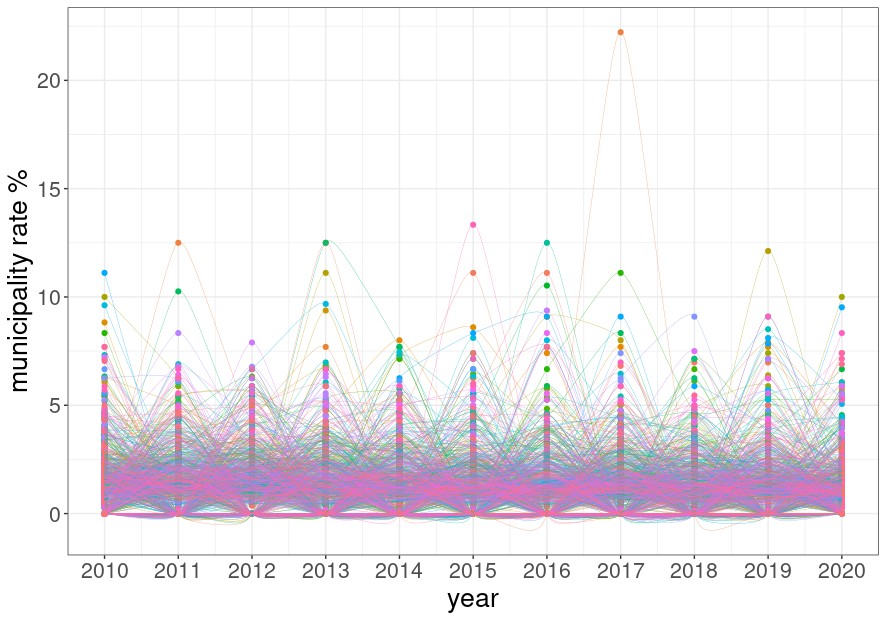}
  %\caption{A subfigure}
  %\label{fig:sub1}
\end{subfigure}%
\begin{subfigure}{.5\textwidth}
  \centering
  \includegraphics[width=\linewidth]{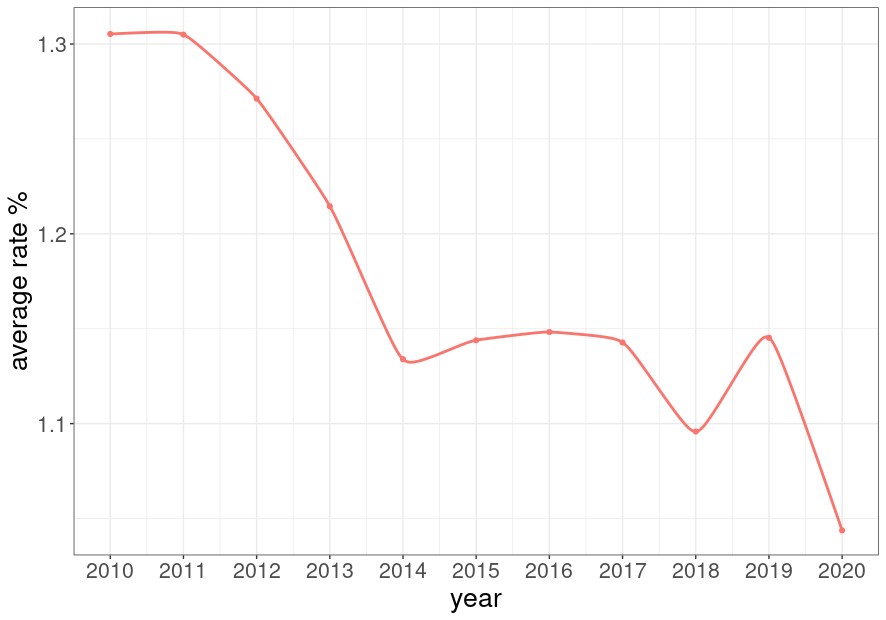}
  %\caption{A subfigure}
  %\label{fig:sub2}
\end{subfigure}
\caption{Infant mortality rate $\%$ in the municipalities and the average rate $\%$ in Minas Gerais, Brazil, from 2010 to 2020}
     \label{fig:trends}
\end{figure}

Infant mortality rates are considered as an important indicator of overall regional development and health conditions. Figure \ref{fig:trends} show an overall decline in the average mortality rate $\%$ in Minas Gerais from 1.31 in 2010 to 1.04 in 2020, and the steepest decline rate is between 2019 and 2020. The observed mortality rates for the states in each area of this region vary each year; see Figure \ref{fig:maps}.

We fit the spatiotemporal variations with the four interaction types presented in Section \ref{interactiontypes1234} using Poisson likelihood,
\begin{center}
    $\pmb y \sim \text{Poisson}(\pmb \vartheta e^{\pmb \eta})$,
\end{center}
where $\pmb y$ is the responses, $\pmb \vartheta$ is the number of expected cases, and $\pmb \eta$ is the linear predictor. We model $\pmb \eta$ by the classical parameterization of \cite{KnorrHeld2000BayesianMO}, as in (\ref{sim1_example}), with RW$2$ and Besag models as the main structured effects. We test the four different interaction types for the interaction term and assign the PC-joint prior \cite{FrancoVilloria2022VariancePI} for the hyperparameter, 

\begin{equation}
    \pi(\pmb \theta) = \dfrac{\lambda b}{4} (1 - \exp(-b))^{-1} \tau_\varepsilon^{1/2} (\textstyle \sum_* \tau_*^{-1})^{2} \exp\Big(- \lambda (\textstyle \sum_* \tau_*^{-1})^{1/2} -b \tau_\varepsilon^{-1/2} (\sum_* \tau_*^{-1})^{-1/2}\Big) \Big |\dfrac{\partial \pmb \vartheta}{\partial \pmb \theta} \Big|,
    \label{pcjointprior}
\end{equation}

where $\lambda$ and $b$ are computed by,
\begin{equation}
    \dfrac{1-\exp(-b\sqrt{U_1})}{1 - \exp(-b)} = a_1 ~~~\text{and} ~~~ \lambda = \dfrac{\log(a_1)}{U_2}.
\end{equation}
The PC-joint prior quantifies the relative contribution of the main and interaction components to the total variance and eases the interpretation of the precision parameters. The parameters $U_1$, $U_2$, $a_1$, and $a_2$ are defined by the user. As a rule of thumb, we use $U_1 = 0.5$, $U_2 = 1/0.31$, $a_1 = 0.99$, and $a_2 = 0.01$ \cite{Simpson2014PenalisingMC, FrancoVilloria2022VariancePI}. The partitions $\pmb \vartheta  = (\vartheta_1,\vartheta_2,\vartheta_3,\vartheta_4,\vartheta_5$) of the variances are parameterized, 
\begin{equation}
\begin{split}
        \tau_\alpha^{-1} &= \vartheta_1^{-1} (1 - \vartheta_2)(1 - \vartheta_3)(1 - \vartheta_4)\\
\tau_\gamma^{-1} &= \vartheta_1^{-1} (1 - \vartheta_2)(1 - \vartheta_3)\vartheta_4\\
\tau_\delta^{-1} &= \vartheta_1^{-1} (1 - \vartheta_2)\vartheta_3(1 - \vartheta_5)\\
\tau_\phi^{-1} &= \vartheta_1^{-1} (1 - \vartheta_2)\vartheta_3 \vartheta_5\\
\tau_\varepsilon^{-1} &= \vartheta_1^{-1} \vartheta_3 (1 - \vartheta_2)\vartheta_5
\end{split},
\end{equation}

with PC priors for $\vartheta_1$, $\vartheta_2$ and uniform priors for $\vartheta_3$, $\vartheta_4$,$\vartheta_5$ \cite{FrancoVilloria2022VariancePI}. We compute the marginal posteriors of the hyperparameters and the latent field for each type using \texttt{inla1234()} from the \texttt{INLAPLUS} package to fit the four interaction types.

%Using \texttt{inla()}, inference takes 6.81 hours, while when using \texttt{inla1234()}, it takes 0.31 hour which is 21 times faster.  

{ 
\renewcommand{\arraystretch}{1.4}
\begin{table}[hbt!]
\centering
\begin{tabular}{c|cc}
\textbf{Type} & \textbf{DIC} & \textbf{logmlik} \\ \hline
\textbf{1}    & 29026        & -14581.7         \\
\textbf{2}    & 29003        & -14580.4         \\
\textbf{3}    & 29026        & -14582.1         \\
\textbf{4}    & 27995        & -14223.4              
\end{tabular}
\caption{Comparing model selection criteria for the four interaction types}
\label{DIClogmlik}
\end{table}
}{ \renewcommand{\arraystretch}{0.5}}

Log-Marginal likelihood (logmlik) and DIC are extensively adopted for model selection criteria.  We report these values for the interaction types \rom{1}, \rom{2}, \rom{3}, and \rom{4} in Table \ref{DIClogmlik}. Comparing the models, interaction type \rom{4} better fits the mortality data in Minas Gerais. 
To investigate the variance contribution of each random effect, we use the posterior mean of the partition parameters $\vartheta_1$, $\vartheta_2$, $\vartheta_3$, $\vartheta_4$, and $\vartheta_5$ and interpret the variabilities, see Table \ref{partition::inter}.

{ 
\renewcommand{\arraystretch}{1.4}
\begin{table}[]
\centering
\resizebox{\columnwidth}{!}{%
\begin{tabular}{|l|l|}
\hline
\textbf{Parameters}                         & \textbf{Interpretation}                                                                                                                                                                  \\ \hline
$\vartheta_1 = 26.404$                      & is the overall precision parameter and $\vartheta_1^{-1} = 0.0372$ is the generalized variance.                                                                                         \\ \hline
$\vartheta_2 =  0.028$                      & is the proportion of the total variance interpreted by the interaction effect                                                                                                            \\ \hline
$\vartheta_1^{-1}(1 - \vartheta_2) = 0.036$ & is the variance explained by the main effects.                                                                                                                                           \\ \hline
$(1 - \vartheta_3) = 0.349$                 & \begin{tabular}[c]{@{}l@{}}variance that quantifies the proportion of variance explained by the main effects\\ and which are attributed to the temporal effects.\end{tabular}               \\ \hline
$(1 - \vartheta_4) = 0.396$                 & \begin{tabular}[c]{@{}l@{}}variance that quantifies the proportion of variance explained by the temporal effects\\ and which are attributed to the structured temporal effects.\end{tabular} \\ \hline
$(1 - \vartheta_5) = 0.755$                 & \begin{tabular}[c]{@{}l@{}}variance that quantifies the proportion of variance explained by the spatial effects\\ and which are attributed to the structured spatial effects.\end{tabular}   \\ \hline
\end{tabular}}
\caption{Variance partitions and their interpretations when using interaction type \rom{4}}
\label{partition::inter}
\end{table}
 }{ \renewcommand{\arraystretch}{0.5}}
 
\begin{figure}[hbt!]
\centering
\begin{subfigure}{.5\textwidth}
  \centering
  \includegraphics[width=\linewidth]{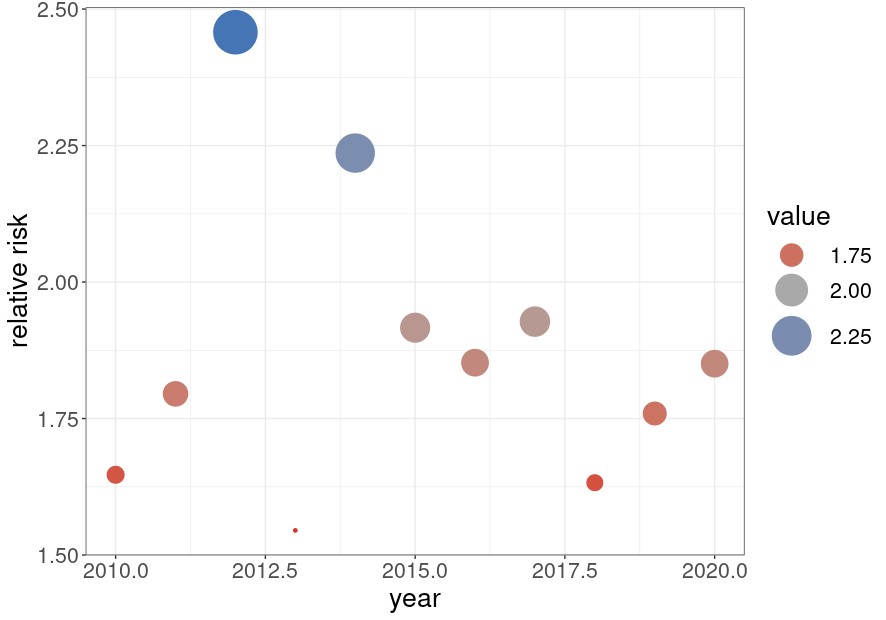}
    \caption{structured and unstructured temporal effects}
  %\label{fig:sub1}
\end{subfigure}%
\begin{subfigure}{.5\textwidth}
  \centering
  \includegraphics[width=\linewidth]{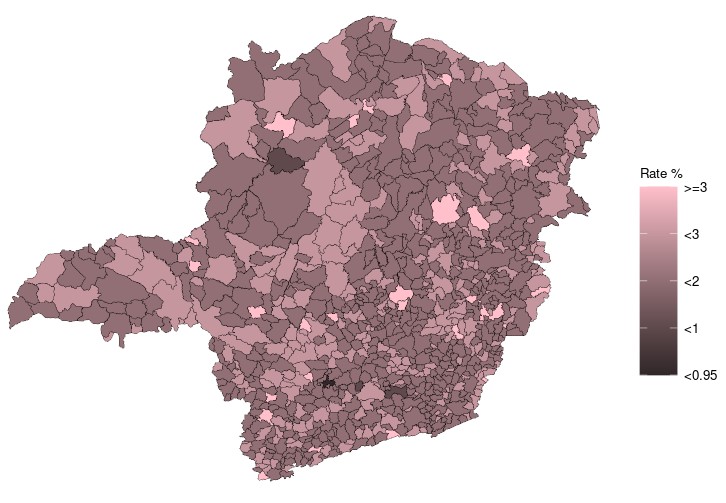}
  \caption{structured and unstructured spatial effects}
  %\label{fig:sub2}
\end{subfigure}
\caption{Posterior Mean of the temporal and spatial relative risk considering interaction type \rom{4}}
     \label{fig:relrisk}
\end{figure}

The high infant mortality in Minas Gerais state cannot be mainly due to an infectious disease as the estimated contribution of the main effects is about $97.1\%$, which means the interaction effect plays a minor role in explaining the variability in mortality risk. The geographic location of municipalities explain $65.08\%$ of the risk variability due the main effects. This highlights the inequalities in infant mortality between different municipalities, which is possibly due to socio-economic factors and the accessibility of medical health centers in various areas. Lastly, structured effects for both space and time are predominant. Figure \ref{fig:relrisk} shows that the relative risks for spatial effects lead to enormous inequalities in infant mortality in Minas Gerais, although the posterior mean of the temporal effects show an overall decay from 2012 to 2020. These results might be helpful for policy makers when deciding which municipalities require more primary healthcare.

%Posterior mean of the spatial relative risk 

%\begin{figure}
%\centering
%\begin{subfigure}{.5\textwidth}
%  \centering
%  \includegraphics[width=\linewidth]{figures/srelative.jpg}
  %\caption{}
  %\label{fig:sub1}
%\end{subfigure}%
%\begin{subfigure}{.5\textwidth}
%  \centering
%  \includegraphics[width=\linewidth]{figures/trelative.jpg}
  %\caption{A subfigure}
  %\label{fig:sub2}
%\end{subfigure}
%\caption{Posterior mean of the spatial and temporal relative risk}
%     \label{fig:trends}
%\end{figure}

\section{Conclusion}

Complexity in disease mapping models for the interaction types \rom{1}, \rom{2}, \rom{3}, and \rom{4} is no longer an upcoming challenge, but an ongoing issue. As a result of the advancement in computational power and increased data collection, alternatives to current methods are necessary in order to illustrate possible associations between diseases and the environment and fit disease-mapping models with high interactions. In this paper, we present a new framework based on INLA methodology to fit disease-mapping models for the interaction types \rom{1}, \rom{2}, \rom{3}, and \rom{4} that bypasses the constraints complexity issue. The addressed approach takes advantage of the improved availability and accessibility of multiple nodes of multi-core architectures on shared and distributed memory to speed-up inference. The framework employs the use of dense structures for precision/covariance matrices to bypass the identifiability issues that arise from adding sum-to-zero constraints. The use of pseudo-inverse ensures inference in the new approach is independent of the complexity of the interactions. This opens the stage to test models with interactions between more than two main effects and may allow epidemiologists, policy makers, and health researchers to obtain inferences for more complex problems.

{9}

\newpage
\appendix
\section*{Appendices}
\addcontentsline{toc}{section}{Appendices}
\renewcommand{\thesubsection}{\Alph{subsection}}

\subsection{Fitting Spatiotemporal with Interaction Types \rom{1}, \rom{2}, \rom{3} and \rom{4} using INLA}
\label{Appendix A}

In this appendix, we review the main key concepts needed as preliminaries for Integrated Nested Laplace Approximation (INLA) methodology.

\subsubsection{Intrinsic Gaussian Markov Random Field (IGMRF)}

An Intrinsic Gaussian Markov Random Field (IGMRF) $\pmb x \in \mathbb{R}^s $ is a multivariate Gaussian distribution, with mean $\pmb \mu$ and a rank deficient matrix $\pmb Q$ of rank $r$, has the density,
\begin{equation}
    \pi(\pmb x) = (2 \pi)^{-s/2} (|\pmb Q|^*)^{1/2} \exp \Big( \frac{1}{2}(\pmb x - \pmb \mu)^T\pmb Q(\pmb x - \pmb \mu)  \Big)
\end{equation}
where $|\pmb Q|^*$ is the generalized determinant of this semi-positive definite matrix. The order of an IGMRF can be defined as the rank deficiency of this precision matrix. The canonical parameterization of $\pmb x$ is defined as a function of $\pmb b$ and $\pmb Q$,
\begin{equation}
    \log \pi(\pmb x) = - \frac{1}{2} \pmb x^T \pmb Q \pmb x + \pmb b^T \pmb x + \text{constant},
\end{equation}
where $\pmb b = \pmb Q \pmb \mu $.

\subsubsection{INLA Methodology for Intrinsic Models} \label{INLAmethodology}

\begin{figure}[hbt!]
\centering
\begin{tikzpicture}
\node[shape=circle,draw=black,thick] (x) at (0.75,3) {$\pmb x$};
\node[shape=circle,draw=black,thick] (tx) at (0,4.4) {$\pmb \theta_x$} ;
\node[shape=circle,draw=black,thick] (ty) at (1.5,4.4) {$\pmb \theta_y$};
\node[shape=rectangle,draw=black,thick,minimum width=0.5cm,minimum height=1cm ] (y) at (1.5,1.5) {$\pmb y$} ;

%\node[shape=circle,draw=black] (mu) at (-1.5,3.5) {$\mu$} ;		   
%\node[shape=circle,draw=black] (mu) at (-1.5,3.5) {$\mu$} ;
%\node[shape=circle,draw=black] (muzer) at (-3,4) {$\mu_{0}$} ;
%\node[shape=circle,draw=black] (taumuzer) at (-1.5,4.9) {$\tau_{\mu}$} ; 
\path [->] (x) edge node[right] {} (y);
\path [->] (ty) edge node[right] {} (y);
 \path [->] (tx) edge node[left] {} (x);

\end{tikzpicture}
\caption{Structure of Latent Gaussian Models in INLA} \label{fig:SLGMINLA}
\end{figure}
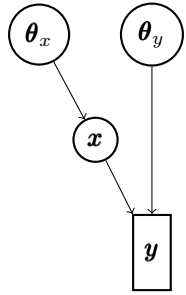

Latent Gaussian Models are an extensively used class of models with a wide range of applications: generalized linear models, generalized additive models, and spatial and spatiotemporal models, amongst others. INLA focuses on these types of models. The structured latent Gaussian models in INLA are defined in terms of three-level hierarchical formulation, the hyperparameter $\pmb \theta = (\pmb \theta_x, \pmb \theta_y)$, the IGMRF $\pmb x | \pmb \theta$, and the likelihood model $ \pmb y|\pmb x, \pmb \theta$, see Figure \ref{fig:SLGMINLA}. 

The main interest of INLA is to approximate the marginal posteriors of the hyperparameters and the latent field elements \cite{Rue2009ApproximateBI}. We summarize the main steps in two sequential stages.

\vspace{0.3cm}
\noindent \textbf{\underline{Stage 1 is to approximate:}}
\begin{equation}
\pi(\theta_{i}|\pmb y) = \displaystyle\int \pi(\pmb \theta|y)~ d\pmb \theta_{-i} \propto \displaystyle\int_{\pmb x} \displaystyle\int_{\pmb \theta_{-i}} \pi(\pmb y|\pmb x,\pmb \theta) \pi(\pmb x|\pmb \theta) \pi(\pmb \theta)  d\pmb \theta_{-i} d\pmb x
\end{equation}
\begin{itemize}
    \item Construct an approximation for the marginal posterior of $\pmb \theta$, $\tilde{\pi}(\pmb \theta|y)$ to find the mode $\pmb \theta^*$.
    \item Capture some of the asymmetries of $\tilde{\pi}(\pmb \theta|y)$ using scaling parameters on each direction of each axis of $\pmb \theta$ \cite{Martins2013BayesianCW}.
    \item Find the marginal posterior $\pi(\pmb \theta_i|\pmb y)$ for each $i$ by integrating out some sequences of points of the hyperparameter using the scaled parameters in the previous step.
\end{itemize}
\vspace{0.2cm}
\noindent \textbf{\underline{Stage 2 is to approximate:}}
\begin{equation}
    \label{marginallatentxapprox}
    \pi(x_{i}|\pmb y) = \displaystyle\int \pi(x_i|\pmb \theta, \pmb y) \pi(\pmb \theta|\pmb y) ~ d\pmb \theta \propto \displaystyle\int_{\pmb x_{-i}} \displaystyle\int_{\pmb \theta} \pi(\pmb y|\pmb x,\pmb \theta) \pi(\pmb x|\pmb \theta) \pi(\pmb \theta)  d\pmb \theta d\pmb x_{-i}
\end{equation}
\begin{itemize}
 \item Approximate the full conditional distribution of the latent field $\tilde{\pi}(x_i |\pmb \theta,\pmb y)$.
    \item Explore a set of $\pmb \theta$s: $\{\pmb \theta_k\}_k$ and their respective weights: $\{\Delta_k\}_k$ like \textit{Grid} or \textit{CCD} strategy \cite{Rue2009ApproximateBI}.
    \item Integrate out the evaluation points of $\pmb \theta$ to get the marginal posterior $\tilde{\pi}(x_{i}|\pmb y)$.
\end{itemize}

\subsubsection{Approximations of $\pi(\pmb \theta|\pmb y)$ and $\pi(\pmb x|\pmb y,\pmb \theta)$}

INLA approach constructs $\tilde{\pi}(\pmb \theta|\pmb y)$ by replacing the full conditional of $\pmb x$: $\pi(\pmb x|\pmb y,\pmb \theta)$ with its Gaussian approximation: $\tilde{\pi}_G(\pmb x|\pmb y,\pmb \theta) \sim \mathcal{N}(\pmb x^*, \pmb \Sigma^*)$, then it writes the Laplace approximation of $\pi(\pmb \theta|\pmb y)$, and evaluate it at the mode $\pmb x^*$,
\begin{equation}
   \tilde{\pi}(\pmb \theta|\pmb y) \propto \displaystyle  \frac{\pi(\pmb \theta) \pi(\pmb x|\pmb \theta) \pi(\pmb y|\pmb x,\pmb \theta)}{\tilde{\pi}_G(\pmb x|\pmb y,\pmb \theta)}\Bigg|_{\pmb x^*}.  
\end{equation}
We get the model configuration $\pmb \theta^*$ using Limited-memory Broyden Fletcher Goldfarb Shanno (L-BFGS) algorithm \cite{10.5555/3112655.3112866}, an iterative method for unconstrained nonlinear minimization problems,
$$ \pmb \theta^*  = \underset{\pmb \theta}{\mathrm{argmax}} \hspace{3mm} -\tilde{\pi}(\pmb \theta|\pmb y).$$ 
The estimated gradient $\nabla \tilde{\pi}(\pmb \theta|\pmb y)$ at each iteration for this algorithm and the estimated Hessian $\nabla^2 \tilde{\pi}(\pmb \theta|\pmb y)$ of $\tilde{\pi}(\pmb \theta|\pmb y)$ at $\pmb \theta^*$ are approximated using numerical differentiation methods boosted by the Smart Gradient and Hessian Techniques \cite{AbdulFattah2022SmartG}. We need to correct for the constraints at every configuration of the hyperparameter $\pmb \theta$. INLA computes uncorrected mean $\pmb x^*_{un}$ and uncorrected covariance $\pmb \Sigma^*_{un}$ using sparse solvers, then it uses kriging technique to find $\pmb x^*$ and $\pmb \Sigma^*$.

\subsubsection{Sum-to-Zero Constraints}
\cite{Rue2005GaussianMR} and \cite{Schrdle2011SpatiotemporalDM} deal with these constraints using kriging technique that is based on the conditional moments from the joint distribution of an IGMRF $\pmb x$ and $\pmb C \pmb x \in \mathbb{R}^s$ that represents the $k$ constraints,

\begin{equation}
\pi(\pmb x, \pmb C \pmb x) \sim \mathcal{N} \left( \begin{pmatrix}
\pmb \mu\\
\pmb C \pmb \mu
\end{pmatrix},
 \begin{pmatrix}
\pmb \Sigma^*_{un} & \pmb \Sigma^*_{un} \pmb C^T\\
\pmb C \pmb \Sigma^*_{un} & \pmb C \pmb \Sigma^*_{un} \pmb C^T.
\end{pmatrix} \right )
\end{equation}
Conditioning on $\pmb C \pmb x = \pmb 0$, we get the needed correction,
\begin{equation}
\pmb x^* = \text{E}(\pmb x|\pmb C \pmb x = \pmb 0) = \pmb x^*_{un} - \pmb \Sigma^*_{un} \pmb C^T (\pmb C \pmb \Sigma^*_{un} \pmb C^T)^{-1} (\pmb C \pmb x^*_{un}),
\end{equation}
and 
\begin{equation}
\pmb \Sigma^* = \text{Cov}(\pmb x|\pmb C \pmb x = \pmb 0) = \pmb \Sigma^*_{un} - \pmb \Sigma^*_{un} \pmb C^T (\pmb C \pmb \Sigma^*_{un} \pmb C^T)^{-1} (\pmb C \pmb \Sigma^*_{un}).
\label{kriggCov}
\end{equation}
The product $\pmb C \pmb \Sigma^*_{un} \pmb C^T$ is a dense structure; hence, its factorization is expensive for large $k$. The additional cost of having $k$ constraints is $\mathcal{O}(sk^2)$. Its computation is intensively required in the two stages to get the inference.

\subsection{\texttt{inla1234()} Function}

We introduce the \texttt{inla1234()} function for the disease-mapping models in the \texttt{INLAPLUS} package. This is a wrapper function that combines INLA and the new approach for these interaction types \rom{1}, \rom{2}, \rom{3}, and \rom{4}. This function runs the code using the approach that fits the application better, and this is determined by the matrix sparsity and the number of constraints needed. For instance, we only need to add 1-3 constraints for interaction type \rom{1}, then \texttt{inla()} function is used. In the case when the ratio of the number of effects over the number of constraints is less than $r$ ($r$ depends on the number of nodes and processors provided by the user), then the code uses the new approach. The new approach only requires the main structured and non-structured effects, and is built-in to construct the interaction structure and any needed constraints for precise inference. There is no need for the practitioners to add any constraint to the \texttt{inla1234()} function. 

The following code can be used to fit the four interaction types \rom{1}, \rom{2}, \rom{3}, and \rom{4} without the need to add any constraints.

\vspace{0.4cm}
\begin{verbatim}

library("INLA")
library("INLAPLUS")
    
observations =
expected_counts = 
struc_time =   #structured temporal structure R
struc_space =  #structured spatial structure R
r1 =           #rank deficient of the structured temporal structure R
r2 =           #rank deficient of the structured spatial structure R
num_nodes =    #number of nodes available on the system
num_threads =  #number of threads on each node
  
myData = list(y_response = observations)
myModel = list(like="Poisson",  offset = expected_counts)
parallel_strategy = list(nodes = num_nodes, threads = num_threads)

blocks =  y ~ block(type="intercept") + 
              block(type="RW1", rankdef=r1, Q = struc_time) + 
              block(type="iid") + 
              block(type="besag1", rankdef= r2, Q = struc_space) + 
              block(type="iid") + 
              block(type="interaction4") 

#Strategy: is the strategy used to approximate the marginal posteriors

results = inla1234(formula = blocks,
                   Model = myModel,
                   control_strategy = list(Strategy = "GA"),
                   data = myData,
                   parallelize = parallel_strategy)
    
\end{verbatim}

For more details on the function, a tutorial on how to fit disease-mapping models using \texttt{INLAPLUS} package and python can be found at \href{https://github.com/esmail-abdulfattah/INLAPLUS}{github.}

\end{document}